\documentclass[aps, prd, amsmath, floats, floatfix, twocolumn, nofootinbib, superscriptaddress, showpacs]{revtex4-1}
\usepackage{graphicx}
\usepackage{multirow}
\usepackage{color}
\usepackage{latexsym}
\usepackage{amsfonts}
\usepackage[colorlinks]{hyperref}
\graphicspath{{figures/}} 

\newcommand{\be}{\begin{eqnarray}}
\newcommand{\ee}{\end{eqnarray}}

\begin{document}

	\title{Spinning test particle motion around a traversable wormhole}

	\author{Carlos~A.~Benavides-Gallego}
	\email[Corresponding author:]{cabenavidesg20@shao.ac.cn}
	\affiliation{Shanghai Astronomical Observatory, 80 Nandan Road, Shanghai 200030, P. R. China}

	\author{Wen-Biao Han}
	\email[Corresponding author:]{wbhan@shao.ac.cn}
	\affiliation{Shanghai Astronomical Observatory, 80 Nandan Road, Shanghai 200030, P. R. China}
	
	\author{Daniele~Malafarina}
	\email{daniele.malafarina@nu.edu.kz}
	\affiliation{Department of Physics, Nazarbayev University, 53 Kabanbay Batyr avenue, 010000 Nur-Sultan, Kazakhstan }
	
	\author{Bobomurat Ahmedov}
	\email{ahmedov@astrin.uz}
    \affiliation{Ulugh Beg Astronomical Institute, Astronomy St. 33, 
		Tashkent 100052, Uzbekistan}
		\affiliation{National University of Uzbekistan, Tashkent 100174, 
		Uzbekistan}
		\affiliation{Tashkent Institute of Irrigation and Agricultural Mechanization Engineers, Kori Niyoziy, 39, Tashkent 100000, Uzbekistan}

	\author{Ahmadjon Abdujabbarov}
	\email{ahmadjon@astrin.uz}
	\affiliation{Shanghai Astronomical Observatory, 80 Nandan Road, Shanghai 
		200030, P. R. China}
	\affiliation{Ulugh Beg Astronomical Institute, Astronomy St. 33, 
		Tashkent 100052, Uzbekistan}
	\affiliation{National University of Uzbekistan, Tashkent 100174, 
		Uzbekistan}
		\affiliation{Tashkent Institute of Irrigation and Agricultural Mechanization Engineers, Kori Niyoziy, 39, Tashkent 100000, Uzbekistan}
	\affiliation{Institute of Nuclear Physics, Tashkent 100214, Uzbekistan}

	\date{\today}

	\begin{abstract}
        The motion of spinning test particles around a traversable wormhole is investigated using the Mathisson-Papapetrous-Dixon equations, which couple the Riemann tensor with the antisymmetric tensor $S^{\alpha\beta}$, related to the spin of the particle. Hence, we study the effective potential, circular orbits, and innermost stable circular orbit (ISCO) of spinning particles. We found that the spin affects significantly the location of the ISCO, 
        in contrast with the motion of non-spinning particles, where the ISCO is the same in both the upper and lower universes. On the other hand, since the dynamical four-momentum and kinematical four-velocity of the spinning particle are not always parallel, we also consider a superluminal bound on the particle’s motion. In the case of circular orbits at the ISCO,  we found that the motion of particles with an adimensional spin parameter lower (greater) than $s=-1.5$ ($1.5$) is forbidden. The spin interaction becomes important for Kerr black hole orbiting super massive wormholes (SMWH).   
	\end{abstract}
	
	\maketitle
	

	\section{Introduction}
	In the last years of his life, Einstein tried to find a theory to unify electromagnetism with general relativity (GR). In contrast to some scientists, Einstein was concerned by the idea of considering material particles as singularities of the fields:``\textit{a singularity brings so much arbitrariness in the theory that it nullifies its laws}''\cite{Einstein:1935tc}. In this sense, it was a fundamental principle for Einstein to remove singularities in a field theory. Hence, in 1935, Einstein and Rosen tried to answer the following question:
		\begin{quotation}
		``\textit{Is an atomistic theory of matter and electricity conceivable which, while excluding singularities in the field, makes use of no other field variables than those of the gravitational field ($g_{\mu\nu}$) and those of the electromagnetic field in the sense of Maxwell (vector potentials)?}''\cite{Einstein:1935tc}.
		\end{quotation} 
	One example of such singularities is the curvature singularity that appears in the well-known Schwarzschild solution~\cite{Schwarzschild:1916uq, Schwarzschild:1916ae}. When solving the field equations for a vacuum, static and spherically symmetric space-time (with and without an electrostatic field), it is clear that singularities at the center of symmetry must emerge. Nevertheless, by reinterpreting the Schwarzschild solution, Einstein and Rosen were able to find a way to avoid the singularity problem. Their idea consisted of treating the physical space as two concurrent sheets, where (neutral and charged) particles are seen as the portions of the space-time that connect the two sheets. In simple words, Einstein and Rosen thought of particles as ``\textit{bridges}''\footnote{It is necessary to point out that the Einstein–Rosen bridge was originally discovered by Flamm (1916) after Schwarzschild published his solution~\cite{G.W.Gibbons2015}.} connecting two regions of the space-time.

	With the paradigm of particles as bridges, Einstein and Rosen tried to explain the atomistic character of matter\footnote{For example, according to this idea, the problem of the nature of electrons and protons, which are described as point-like particles in electrodynamics, can be addressed by considering it as a two bridge problem.} without introducing new variables other than the metric and the vector potential. It is worth noticing that for some time the idea of the existence of ``\textit{bridges}'', or  wormholes\footnote{The term ``\textit{wormhole}'' was coined by Wheeler and Misner in 1957~\cite{Misner:1957mt}.} as they are called today, was considered more appealing than that of black holes. However, in 1962, Wheeler and Fuller showed that the Einstein-Rosen bridge (also known as Schwarzschild wormhole) is unstable and will pinch off in a finite time. Hence, although the Schwarzschild space-time is static, a proper analysis shows that the Schwarzschild geometry is changing with time~\cite{Fuller:1962zza}. As a consequence, Schwarzschild wormholes are not traversable.

	The possibility of traversable wormholes in general relativity appeared for the first time in the works of Ellis~\cite{HGEllis:1973} and Bronnikov~\cite{Bronnikov:1973fh} (1973). In his paper, Ellis obtained a solution of Einstein's field equations for a vacuum space-time, modified by the inclusion of a scalar field coupled to the Ricci tensor. The solution is known as the Ellis drain-hole. Thus, by analyzing the topology and geodesics, Ellis was able to show that it is geodesically complete, horizonless, singularity-free, and fully traversable in both directions. The drain-hole solution requires two parameters to be described: $m$, which fixes the strength of its gravitational field, and $n$, which determines the curvature of its spatial cross-sections. When $m=0$, the drain-hole gravitational field vanishes, forming a non-gravitating, purely geometric, traversable wormhole~\cite{HGEllis:1973}. 

	Traversable wormholes motivated numerous studies during the '80s and the '90s ~\cite{Morris:1988tu, Morris:1988cz, Visser:1989kh, Visser:1989kg, Visser:1989vq, AzregAinou:1989wr, Visser:1989am, Poisson:1989zz, Visser:1990wj, Frolov:1990si, Visser:1990wi}. These works provided a deeper understanding of the physics behind each particular solution. In Ref.~\cite{Morris:1988cz}, for example, Morris and Thorne considered the following question: \textit{What properties a classical wormhole should have to be traversable?} Usually, as Visser pointed out in Ref.~\cite{Visser:1989kh}, the process to analyze classical wormholes starts by considering a Lagrangian, solving the Einstein field equations, and then looking for different geometries among these solutions. However, the authors of Ref.~\cite{Morris:1988cz} used a different approach. There, Morris and Thorne assumed the wormhole geometry to derive, via Einstein field equations, the stress-energy tensor and then investigate the physics. In this way, they were able to show that ``\textit{exotic}" matter should be present at the throat of a traversable wormhole. 

	On the other hand, the idea of wormholes also has inspired the possibility of time travel~\cite{Morris:1988cz, Morris:1988tu}. Although this possibility has been considered speculatively (because the existence of wormholes is constrained to existence of ``\textit{exotic}'' matter, which violates the weak energy condition), we cannot entirely dismiss it. In fact, the discovery that black holes can evaporate~\cite{Hawking:1974rv} has suggested that quantum fields can violate the energy conditions and has led some to speculate on the validity of energy conditions~\cite{Martin-Moruno:2013}. 
	One example related to wormholes is the quantum creation of particles in~\cite{Morris:1988cz}.

	After the work of Einstein and Rosen (1935)~\cite{Einstein:1935tc}, the interest in wormhole solutions remained dormant for almost twenty years. Then, with the works of Wheeler (1955) and Misner (1957)~\cite{Wheeler:1955zz, Misner:1957mt} wormholes started to be considered again by the scientific community as viable astrophysical objects. Recently, a lot of theoretical researchers have studied different aspects of traversable wormholes within Einstein's gravity~\cite{Echeverria:1991nk,Deser:1992ts,Deser:1993jr,Hochberg:1997wp,Abdujabbarov:2009ad,Abdujabbarov:2016efm} as well as in alternative theories of gravity~\cite{Moffat:1991xp,Carlini:1992jda,Bhawal:1992sz,Letelier:1993cj,Vollick:1998qf}. Furthermore, some authors have considered the effects that wormholes would produce from the observational point of view~\cite{Perlick:2003vg, Perlick:2004tq, Nandi:2006ds, Kardashev:2006nj, Bambi:2013jda, Tripathi:2019trz} with the aim of putting constraints on their possible detection. In~\cite{Bambi:2013jda}, for example, it is studied the K$\alpha$ iron line of several wormhole solutions based on the idea that supermassive black hole candidates at the center of galaxies might be wormholes formed in the early universe. By calculating the K$\alpha$ iron line produced by accretion disks in the space-time of these solutions, one could compare it with that produced by a Kerr black hole. It was found that the K$\alpha$ iron line produced around non-rotating or slow-rotating wormholes may mimic the one obtained in vicinity of Kerr black holes (with mid or high spins). Moreover, the results are still marginally compatible with current observations. Hence, the possibility that the supermassive black hole candidates in galactic nuclei could in fact be these objects is still not ruled out. On the other hand, in the case of wormholes with spin parameter $a_{*}>0.02$, the iron line is indeed different from the one produced in the space-time of a Kerr black hole and therefore, their existence may already be excluded via current observations~\cite{Bambi:2013jda}. 
		 
    One aspect that is important to consider when dealing with accretion disks is the spin of the particles in the accretion disk's gas. In fact nowadays, the spin interaction of relativistic systems has become an important subject of study~\cite{Kidder:1992fr,Apostolatos:1994mx, Suzuki:1996gm,Suzuki:1997by,Saijo:1998mn,Semerak:1999qc,Kyrian:2007zz,Plyatsko:2013xza,Hackmann:2014tga,Nucamendi:2019qsn,Conde:2019juj,Toshmatov:2019bda,Boonserm:2019nqq,Han:2016cdh,Han:2016djt,Toshmatov:2020wky}. For this reason, it is important to properly understand the dynamics of test particles in curved space-time that includes classical spin. In 1937, Mathisson studied the problem of extended bodies in GR. In his work, he demonstrated an existing interaction between the Riemann curvature tensor and the spin of the moving particle in the equations of motion~\cite{Mathisson:1937zz}. Papapetrou also considered the same problem in Refs.~\cite{Papapetrou:1951pa, Corinaldesi:1951pb} developing a similar approach as Mathisson. It is worth remembering that later, Tulczyjew improved on the methods of Mathisson~\cite{tulczyjew1959motion, BWTulzcyjew1962} while Moller and others made improvements in the definition of center-of-mass in Refs.~\cite{moller1949definition,beiglbock1967center,dixon1964covariant, Dixon:1970zza,Dixon:1970zz,ehlers1977dynamics}. Today, the equations that describe the motion of extended bodies with spin and mass are known as the Mathisson-Papapetrous-Dixon equations (MPD). 
  	
    In this work, we investigate the motion of spinning particles around a traversable wormhole. We organize our work as follows: In Sec.~\ref{SecI}, we review properties of traversable wormholes discussed in Ref.~\cite{Morris:1988cz}. Next, in Sec.~\ref{SecII}, we discuss the motion of spinning particles in a spherically symmetric space-time. In this section, we use the MPD equations to obtain the effective potential and the superluminal condition. Then, in Sec.~\ref{SecIII} and Sec.~\ref{Section IV}, we apply the results of Sec.~\ref{SecII} to the Morris-Thorne wormhole. We compute the effective potential, circular orbits, and the innermost stable circular orbit (ISCO). We also use the superluminal bound to find a constraint for the allowed spin of the particle. Finally, in Sec.~\ref{SecV}, we summarize our work and discuss the implications of the results. Throughout the manuscript, we use geometrized units setting $G=c=M=1$.
 	
	\section{Morris-Thorne wormholes \label{SecI}}
	
	In this section, we review properties of wormholes discussed by Morris and Thorne in Ref.~\cite{Morris:1988cz}, where the authors assumed the wormhole's space-time geometry, and then, via the field equations, computed the corresponding energy-momentum tensor.
	Therefore, the discussion starts by assuming a static and spherically symmetric space-time for which the line element has the following form 
	\begin{equation}
	\label{2.1}
	ds^2=-e^{2\Phi(r)}dt^2+\frac{dr^2}{\left(1-\frac{b(r)}{r}\right)}+r^2(d\theta^2+\sin^2\theta d\varphi^2),
	\end{equation}
	where $\Phi(r)$ and $b(r)$ are arbitrary functions of the radial coordinate $r$ known as the ``\textit{redshift function}'' and the ``\textit{shape function}'', respectively. The fact that Eq.~(\ref{2.1}) can represent a wormhole with a throat is easily explained by embedding the line element in a three-dimensional space at a fixed time slice $t$. In addition, given the spherical symmetry of the line element in Eq.~(\ref{2.1}), we can limit the attention to the equatorial plane $\theta=\pi/2$ and consider the restricted line element  
	\begin{equation}
	\label{2.2}
	ds^2=\left(1-\frac{b(r)}{r}\right)^{-1}dr^2+r^2d\varphi^2.
	\end{equation} 
	\begin{figure}
		\begin{center}
		\includegraphics[scale=0.35]{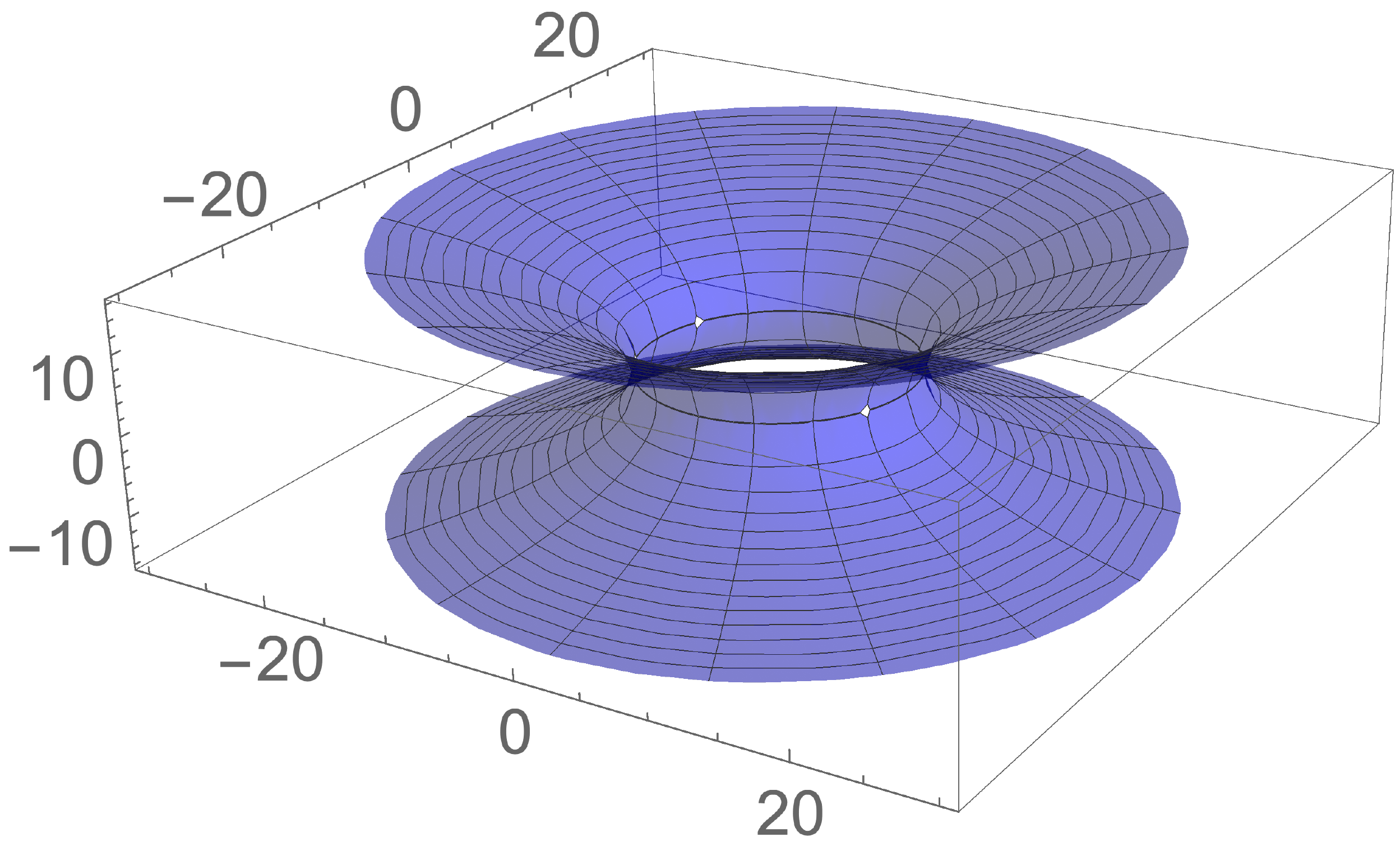}
		\caption{Representation of the wormhole embedded in a three dimensional space-time. For the plot we use the line element $ds^2=-dt^2+dl^2+(b^2_0+l^2)(d\theta^2+\sin^2\theta d\varphi^2)$, which is a special case of Eq.~(\ref{2.1}), where the coordinate $l$ is related to $r$ by $r^2=b^2_0+l^2$. Here the ``\textit{redshift function}'' vanishes ($\Phi=0$) and the ``\textit{shape function}'' is given by $b(r)=b^2_0/r$. In the plot we chose $b_0=10M$.\label{fig2.2}}
		\end{center}
	\end{figure}
	One can interpret Eq.~(\ref{2.2}) as a ``\textit{picture}'' of the space-time observed from the equatorial plane at time $t$. The line element in Eq.~(\ref{2.2}) can be embedded in the three-dimensional space in cylindrical coordinates $\{r,z,\varphi\}$ with $z=z(r)$ as
	\begin{equation}
	\label{2.3}
	ds^2=\left[1+\left(\frac{dz}{dr}\right)^2\right]dr^2+r^2d\varphi^2,
	\end{equation}
	if we compare Eqs.~(\ref{2.2}) and (\ref{2.3}) to obtain $dz/dr$, which is given by
	\begin{equation}
	\label{2.4a}
	\frac{dz}{dr}=\pm \left[\frac{r}{b(r)}-1\right]^{-\frac{1}{2}}.
	\end{equation}

\begin{figure}[t]
		\begin{center}
			\includegraphics[scale=0.3]{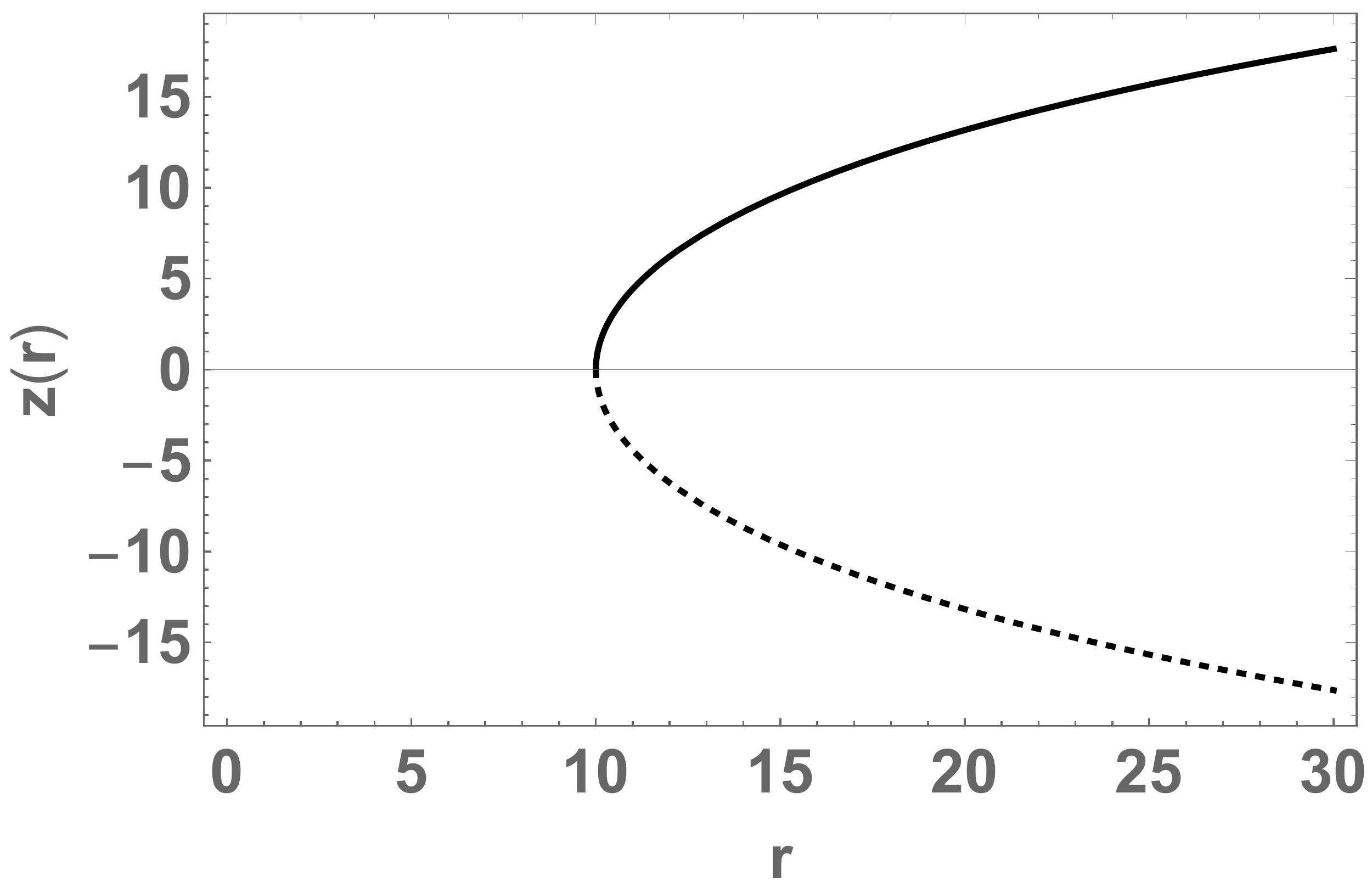}
			\caption{Plot of Eq.~(\ref{2.4}) with $b_0=10M$ (the radius of the throat). One can see the \textit{upper} (continous line) and \textit{lower} (dotted line) universes.  \label{fig2.1}}
		\end{center}
	\end{figure}

	As an example, we consider one of the wormhole solutions obtained in Ref.~\cite{Morris:1988cz} with $b(r)=b_0^2/r$ (see Fig.~\ref{fig2.2} for details). Hence, after integration, the coordinate $z$ as a function of radial coordinate $r$ takes the following form
	\begin{equation}
	\label{2.4}
	z(r)=\pm b_0\ln\left[\frac{r}{b_0}+\sqrt{\left(\frac{r}{b_0}\right)^2-1}\right].
	\end{equation}
	Note that $dz/dr$ in Eq~(\ref{2.4a}) (the slope) diverges at $r=b(r)=b_0$. Moreover, one can observe that Eq.~(\ref{2.4}) has two branches: one with positive sign and the other with negative sign. In  Fig.~\ref{fig2.1} we show the plot of $z$ vs. $r$. From the figure, it is possible to see the two branches of Eq.~(\ref{2.4}), which correspond to the \textit{upper} ($+$) and \textit{lower} ($-$) universes. Here the value $b_0$ is the wormhole's throat radius, where $dz/dr$ diverges. For a complete three-dimensional view of the wormhole, it is necessary to rotate the function $z(r)$ around the $z$-axis. The result is shown in Fig.~\ref{fig2.2}. 

	Now, knowing that the line element in Eq.~(\ref{fig2.1}) does represent a spherically symmetric wormhole, the next step is to make sure it is a solution of the field equations. In this sense, it is necessary to compute the 
    energy-momentum tensor via Einstein's field equations. Here, the reason for considering an energy-momentum tensor different from zero has to do with the fact that while non-traversable wormholes, like the Einstein-Rosen bridge, are vacuum solutions of the field equations for traversable wormholes one must have an energy-momentum tensor different from that of a vacuum located at the throat in order to allow the matching of the upper and lower universes.
		
	The analysis presented in Ref.~\cite{Morris:1988cz} is performed using a ``\textit{proper reference frame}'', this means that physical ``observations" are performed by a local observer, who remains at rest with respect to the coordinate system $\{t,r,\theta,\varphi\}$. The ``\textit{proper reference frame}'' is constructed using the tetrad formalism with the basis vectors~\cite{Chandrasekhar:1998}
		\begin{equation}
	\label{2.5}
	\begin{array}{ccc}
	\textbf{e}_{(a)}=e^{\;\;\;\; \mu}_{(a)}\partial_\mu&\text{and}&	\textbf{e}^{(a)}=e^{(a)}_{\;\;\;\; \mu}dx^\mu.
	\end{array}
	\end{equation}
	Here we use the same notation as in  Ref.~\cite{Chandrasekhar:1998}, where tetrad indices are enclosed in parentheses to distinguish from the tensor indices. In the case of the line element described by Eq.~(\ref{2.1}), the orthonormal basis of a ``\textit{proper reference frame}'' is given by $e^{\;\;\;\; \mu}_{(a)}$. In terms of the metric, the tetrad has the form
	\begin{equation}
	\label{2.6}
	\begin{aligned}
	e^{\;\;\;\;\mu}_{(0)}&=(e^{-\Phi},0,0,0)\\
	e^{\;\;\;\;\mu}_{(1)}&=\left(0,\left[1-\frac{b(r)}{r}\right]^{1/2},0,0\right)\\
	e^{\;\;\;\;\mu}_{(2)}&=(0,0,r^{-1},0)\\
	e^{\;\;\;\;\mu}_{(3)}&=(0,0,0,(r\sin\theta)^{-1}).\\
	\end{aligned}
	\end{equation} 
	The purpose of using a ``\textit{proper reference frame}'' (or, equivalently, a ``\textit{local Lorentz frame}'') in the analysis of the physical quantities is to diagonalize the energy momentum tensor $T_{(a)(b)}$ and then be able to relate each (diagonal) component to a physical quantity of clear meaning, e. i. the total density of mass-energy $\rho(r)$, the tension per unit area $\tau(r)$, and the pressure $p(r)$. Hence, from the physical point of view, the observer in a ``\textit{local Lorentz frame}'' interprets $\rho(r)$, $\tau(r)$, and $p(r)$ as
	\begin{equation}
	\label{2.8}
	\begin{array}{ccc}
	T_{(0)(0)}=\rho ,&T_{(1)(1)}=-\tau(r),&T_{(2)(2)}=T_{(3)(3)}=p(r).
	\end{array}
	\end{equation} 
	
	The field equations of course are given by 
	\begin{equation}
	\label{2.7}
	G_{(a)(b)}=8\pi T_{(a)(b)},
	\end{equation} 
	where $G_{(a)(b)}$ is the Einstein tensor. 
 
	Using Eq.~(\ref{2.8}) and the field equations~(\ref{2.7}) 
	Morris and Thorne obtained the following relations between the physical quantities $\rho(r)$, $\tau(r)$, $p(r)$ and the metric quantities $\Phi(r)$ and $b(r)$~\cite{Morris:1988cz} 
	 \begin{equation}
	 \label{2.9}
	 \begin{aligned}
	 \rho&=\frac{b'(r)}{8\pi r^2}\\\\
	 \tau&=\frac{b(r)/r-2(r-b(r))\Phi'}{8\pi r^2}\\\\
	 p&=\frac{r}{2}\left[(\rho-\tau)\Phi'-\tau'\right]-\tau,
	 \end{aligned}
	 \end{equation} 
	where ${f}'$ denotes the partial derivative of $f$ with respect to $r$.
	Therefore, solving the field equations in the case of a wormhole corresponds to solving the set of equations~(\ref{2.9}). Notice that these are three equations in five unknown quantities. Accordingly, Morris and Thorne proposed to use this freedom to control the functions $\Phi(r)$ and $b(r)$ (related to the wormhole's geometry) in such a way that the solution generates a traversable wormhole. 

	According to Ref.~\cite{Morris:1988cz}, there are several properties that must be satisfied for the geometry to describe a traversable wormhole. 
	As already mentioned, the assumption that the space-time be static and spherically symmetric greatly simplifies the calculations (for stationary wormholes see for example \cite{stationary-wormhole}, while for static but non spherically symmetric wormholes see \cite{axial-wormhole1, axial-wormhole2, axial-wormhole3}). Secondly, any wormhole solution should contain a throat that connects two asymptotically flat regions of space-time. This property is clearly shown in Figs.~\ref{fig2.2} and~\ref{fig2.1} for a particular form of the shape function $b(r)$. As said, the existence of a throat is related to the divergence of $dz/dr$ at $r=b(r)=b_0$, while to demonstrate that the solution is asymptotically flat, noticing that the radial coordinate is ill-define at the throat, Morris and Thorne considered the \textit{proper radial distance}~\cite{Morris:1988cz}  
	\begin{equation}
	\label{2.10}
	l(r)=\pm \int^r_{b_0}\left[1-\frac{b(\tilde{r})}{\tilde{r}}\right]^{-\frac{1}{2}}d\tilde{r},
	\end{equation} 
	which is finite everywhere if $1-b(r)/r\ge 0$. Using the \textit{proprer radial distance}, Morris and Thorne showed that $dz/dl\rightarrow 0$ as $l\rightarrow\pm\infty$, e.i. the space-time is asymptotically flat.

	Finally, the solution must not contain a horizon. Wormholes allow to causally connect two different portions of the space-time by the throat. Therefore, the presence of a horizon would prevent the two universes to be causally connected. This condition is satisfied by demanding $\Phi(r)$ to be finite everywhere. 

	These properties, along with the field equations~(\ref{2.9}), are the ``\textit{basic wormhole criteria}''. Nevertheless, the authors have also discussed the possibility of tuning the wormhole's parameters to make it possible for humans to travel through them; this requires the crossing-time to be finite for any observer and the tidal force felt by the astronaut to be small. On the other hand, one of the most important results is the necessity of ``\textit{exotic matter}'' to generate the wormhole's space-time curvature. This conclusion comes from the fact that in the throat
	\begin{equation}
	\label{2.11}
	\tau_0>\rho_0.
	\end{equation}
	Physically, this means that the tension $\tau_0$ in the throat must be so large as to exceed the total density mass-energy $\rho_0$, which implies that an observer passing through the throat, with a radial velocity close to the speed of light, would perceives a negative $\tilde{T}_{(0)(0)}$~\cite{Morris:1988cz}
	\begin{equation}
	\label{2.12}
	\tilde{T}_{(0)(0)}=\gamma^2[\rho_0-\tau_0]+\tau_0.
	\end{equation}

	 Physicists have established a series of energy conditions to ensure the physical viability of $T_{\mu\nu}$, such as for example avoiding negative energy densities. However, energy conditions are considered to be valid for classical matter fields but may be violated at quantum level~\cite{Visser:1989am}. Therefore, while Eq.~(\ref{2.11}) violates the weak energy condition, from a quantum perspective, there exist some situations in which such violation may be physically valid, such as, for example, the quantum mechanical creation of particles~\cite{Martin-Moruno:2013sfa}. In this sense one can not entirely rule out the possibility of the existence of the exotic material required for the throat of a traversable wormhole to hold~\cite{Morris:1988cz}. 
	 
	 Finally, with the help of Eqs.~(\ref{2.9}), Morris and Thorne established a method to obtain traversable wormhole solutions. The method is as follows. First, given $\Phi(r)$ and $b(r)$ satisfying the conditions for a traversable wormhole, one can use the first relation in Eq.~(\ref{2.9}) to obtain $\rho(r)$. Then, using the second relation in Eq.~(\ref{2.9}) one can obtain $\tau(r)$. Finally, with $\rho(r)$ and $\tau(r)$, one can find $p(r)$. Following this method, Morris and Thorne obtained three solutions: the zero-tidal-force wormhole, a solution with a finite radial cutoff of the stress-energy, and a solution with exotic matter limited to the throat's vicinity. In this paper, we will focus our attention in the zero-tidal-force solution with 
	 \begin{equation}\label{Phi}
	 \Phi(r)=-\frac{b_0}{r} \;\; \text{and} \;\;  b(r)=\frac{b^2_0}{r}. 
	 \end{equation}
    \section{Equations of motion for a spinning particle \label{SecII}}
    The equations of motion of spinning particles are given by the MPD equations that can be expressed as
	\begin{equation}
	\label{3.1}
	\begin{aligned}
	\frac{Dp^\alpha}{d\lambda}&=-\frac{1}{2}R^\alpha_{\;\;\beta\delta\sigma}u^\beta S^{\delta\sigma},\\
	\frac{D S^{\alpha\beta}}{d\lambda}&=p^\alpha u^\beta-p^\beta u^\alpha,
	\end{aligned}
	\end{equation}
	where $D/d\lambda\equiv u^\alpha\nabla_\alpha$ is the projection of the covariant derivative along the particle's trajectory, $u^\mu=dx^\mu/d\lambda$ is the 4-velocity of the test particle, $p^\alpha$ is the canonical 4-momentum, $R^\alpha_{\;\;\beta\delta\sigma}$ is the Riemann curvature tensor, and $\lambda$ is an affine parameter. The second rank tensor $S^{\alpha\beta}$ is antisymmetric\footnote{Note that for any antisymmetric tensor, the diagonal components vanishe identically, i.e. $S^{\alpha\beta}=0$ if $\alpha=\beta$.}, $S^{\alpha\beta}=-S^{\beta\alpha}$. In general relativity, the geodesic equation is given by  
	\begin{equation}
	\label{3.4}
	\partial_\beta u^\alpha u^\beta+\Gamma^\alpha_{\;\;\;\sigma\beta}u^\sigma u^\beta=0,
	\end{equation}
	which can be expressed in terms of the 4-momentum of the particle and the projection of the covariant derivative along the particle's trajectory as
	\begin{equation}
	\label{3.5}
	\frac{Dp^\alpha}{d\lambda}=0.
	\end{equation} 
	Therefore, if one compares Eqs.~(\ref{3.1}) and (\ref{3.5}), one concludes that spinning particles do not follow a geodesic due to the interaction between the Riemann curvature tensor and the antisymmetric tensor $S^{\alpha\beta}$. 
	
	To solve the MPD, it is necessary to fix the center of mass of the spinning particle by including the condition~\cite{tulczyjew1959motion,Saijo:1998mn}
	\begin{equation}
	\label{3.6}
	S^{\alpha\beta}p_\alpha=0,
	\end{equation} 
	which is the so-called Tulczyjew Spin Supplementary Condition (SSC)~\cite{Saijo:1998mn}. From Eq.~(\ref{3.6}), it turns out that the canonical momentum and the spin of the particle provide two independent conserved quantities given by the relations
	\begin{equation}
	\label{3.7}
	\begin{aligned}
	p^\alpha p_\alpha&=-m^2,\\
	s^2&=\frac{1}{2}S^{\alpha\beta} S_{\alpha\beta}.
	\end{aligned}
	\end{equation}
	However, although the canonical momentum of the spinning particle is conserved, it is important to point out that the squared velocity does not necessarily satisfy the condition
	\begin{equation}
	\label{3.8}
	u_\alpha u^\alpha=-1,
	\end{equation}
	because the 4-vectors $p^\alpha$ and $u^\alpha$ are not always parallel. Therefore, it is necessary to impose an additional condition known as \textit{the superluminal bound} to ensure that the particle's 4-velocity is always smaller than the speed of light. 

	In addition to the conserved quantities resulting from the Tulczyjew-SSC condition, there exist also the conserved quantities associated to the space-time symmetries given by the Killing vectors $\xi^\mu$, which can be expressed as 
	\begin{equation}
	\label{3.9}
	p^\alpha \xi_\alpha-\frac{1}{2}S^{\alpha\beta}\nabla_\beta\xi_\alpha=p^\alpha \xi_\alpha-\frac{1}{2}S^{\alpha\beta}\partial_\beta\xi_\alpha=\text{constant}
	\end{equation} 
	where we have used the fact that the term $S^{\alpha\beta}\Gamma^\gamma_{\;\;\beta\alpha}$ in the convariant derivative vanishes because $S^{\alpha\beta}$ is antisymmetric while  $\Gamma^\gamma_{\;\;\beta\alpha}$ is symmetric.
	
	\subsection{Effective potential\label{SecIIA}}
	
	As mentioned before, the line element of a spherically symmetric wormhole is given by Eq.~\eqref{2.1}. Here, we assume $b(r)=b^2_0/r$, where $b_0$ is the wormhole's throat. In general, $b_0$ is interpreted as the mass of the wormhole in the Newtonian limit (see for example Ref.~\cite{Li:2014coa}). Also we make a common choice for the \textit{redshift function}~\cite{Li:2014coa,Harko:2008vy} taking Eq.~\eqref{Phi}. We base this choice on the behavior of circular orbits around the wormhole, which are unstable when $\partial\Phi/\partial r <0$ and stable when $\partial\Phi/\partial r >0$ (see Ref.~\cite{Harko:2008vy} for details). 

    Using the proper distance $l$, related to $r$ by the relation $r^2=b^2_0+l^2$, as a new radial coordinate, the line element Eq.~\eqref{2.1} takes the form:
    \begin{equation}
    \label{3.11a}
    ds^2=-e^{2\Phi(l)}dt^2+dl^2+(b^2_0+l^2)(d\theta^2+\sin^2\theta d\varphi^2),
    \end{equation}
    with
    \begin{equation}
    \label{3.11b}
    \Phi(l)=-\frac{b_0}{\sqrt{b^2_0+l^2}}.
    \end{equation}
    With this choice the throat of the wormhole is located at $l=0$ and we are then able to consider the lower universe ($l<0$) and the upper universe ($l>0$) in a unified manner. 
    Since the space-time is static and spherically symmetric, the line element allows two Killing vector fields given by 
	\begin{equation}
	\label{10.9}
	\begin{array}{ccc}
    \xi^\alpha=\delta^\alpha_t,&&\xi^\alpha=\delta^\alpha_\varphi,
	\end{array}
	\end{equation}
	which correspond to time translation and rotations, and have associated two conserved quantities, i.e. the energy $E$ and the angular momentum $L$. Using Eq.~(\ref{10.9}), the conserved quantities can be expressed as~\cite{Semerak:1999qc, Conde:2019juj, Toshmatov:2019bda}
	\begin{equation}
	\label{3.11}
	\begin{aligned}
	-E=&p_t-\frac{1}{2}g_{t\alpha ,\beta}S^{\alpha\beta}=p_t-\frac{1}{2}g_{tt,l}S^{tl},\\
	L=&p_\varphi+\frac{1}{2}g_{\varphi\alpha,\beta}S^{\beta\alpha}=p_{\varphi}+\frac{1}{2}g_{\varphi \varphi,l}S^{l\varphi}.
	\end{aligned}
	\end{equation}
	Given the spherical symmetry of the geometry we can restrict the attention to a plane of constant $\theta$, such as the equatorial plane $\theta=\pi/2$. Then $p^\theta=0$ and the metric functions on the equatorial plane depend only on the radial coordinate $l$. Therefore, since $S^{\theta\alpha}=0$, the number of independent components of $S^{\alpha\beta}$ reduces to three. Using Eq.~(\ref{3.6}) one obtains the following relations:
	\begin{equation}
	\label{3.14}
	\begin{aligned}
	S^{t\varphi}&=\frac{p_l}{p_\varphi}S^{lt}=-\frac{p_l}{p_\varphi}S^{tl},\\
	S^{l\varphi}&=-\frac{p_t}{p_l}S^{t\varphi}=\frac{p_t}{p_\varphi}S^{tl}.
	\end{aligned}
	\end{equation}
	Now, from the normalization condition and conservation of angular momentum in Eqs.~(\ref{3.7}) we have 
	\begin{equation}
	\label{3.15}
	\begin{aligned}
	p^2_l&=-g_{ll}\left[g^{tt}p^2_t+g^{\varphi\varphi}p^2_\varphi+m^2\right], \\
	2S^2&=S^{\alpha\beta}S_{\alpha\beta}=2S^{lt}\left(S_{tl}-\frac{p_l}{p_\varphi}S_{t\varphi}+\frac{p_t}{p_\varphi}S_{l\varphi}\right),
	\end{aligned}
	\end{equation}
	and
	\begin{equation}
	\label{3.17}
	\begin{aligned}
	S_{tl}&=g_{t\rho}g_{l\epsilon}S^{\rho\epsilon}=g_{tt}g_{ll}S^{tl},
	\\
	S_{t\varphi}&=g_{t\gamma}g_{\varphi\sigma}S^{\gamma\sigma}=-g_{\varphi\varphi}g_{tt}\frac{p_l}{p_\varphi}S^{tl},
	\\
	S_{l\varphi}&=g_{l\lambda}g_{\mu\varphi}S^{\lambda\mu}=g_{ll}g_{\varphi\varphi}\frac{p_t}{p_\varphi}S^{tl}.
	\end{aligned}
	\end{equation}
	From which we get
	\begin{equation}
	\label{3.20}
	S^{tl}=\pm\frac{p_\varphi s}{\sqrt{-g_{tt}g_{ll}g_{\varphi\varphi}}}= \frac{p_\varphi e^{-\Phi(l)}}{\sqrt{b^2_0+l^2}}s,
	\end{equation}
	where $s=S/m$ can be positive or negative and represents the direction of the spin with respect to the direction of $p_\varphi$. Now, from the conservation of energy and angular momentum Eq.~(\ref{3.11}) we have 
	\begin{equation}
	\label{3.21}
	\begin{aligned}
	-E&=p_t-s\mathcal{A}p_\varphi, \\
	L&=p_\varphi+s\mathcal{B}p_t,
	\end{aligned}
	\end{equation}
	with\footnote{Where now prime ${}'$ denotes partial derivatives with respect to radial coordinate $l$.}

	\begin{equation}
	\label{3.22}
	\begin{aligned}
	\mathcal{A}&=\frac{1}{2}\frac{g_{tt,l}}{\sqrt{-g_{tt}g_{ll}g_{\varphi\varphi}}}=-\frac{ e^{\Phi(l)}\Phi'(l)}{\sqrt{b^2_0+l^2}},\\\\
	\mathcal{B}&=\frac{1}{2}\frac{g_{\varphi\varphi,l}}{\sqrt{-g_{tt}g_{ll}g_{\varphi\varphi}}}=\frac{l e^{-\Phi(l)}}{\sqrt{b^2_0+l^2}}.
	\end{aligned}
	\end{equation}
	After solving, we obtain 
	\begin{equation}
	\label{3.23}
	\begin{aligned}
	p_t&=\frac{-E+s\mathcal{A}L }{1+s^2\mathcal{A} \mathcal{B}},\\\\
	p_\phi&=\frac{L+s\mathcal{B}E}{1+s^2\mathcal{AB}},
	\end{aligned}
	\end{equation}
	and
	\begin{equation}
	\label{3.24}
	\mathcal{AB}=-\frac{l\Phi'(l)}{b^2_0+l^2}.
	\end{equation}
	Now, after inserting Eq.~(\ref{3.23}) into Eq.~(\ref{3.15}), we finally obtain 
	\begin{equation}
	\label{3.25}
	(p_l)^2=\frac{1}{\alpha}(\beta E^2 + 2\delta L E + \sigma L^2 - \gamma ),
	\end{equation}
	which is quadratic equation for $E$ with
	\begin{equation}
	\label{3.26}
	\begin{aligned}
	\alpha&=g^{ll}\left(1-\frac{g'_{tt}g'_{\varphi\varphi} s^2}{4g_{tt}g_{ll}g_{\varphi\varphi}}\right)^2,\\\\
	\beta&=-g^{tt}+\frac{g^{\varphi\varphi}(g'_{\varphi\varphi})^2s^2}{4g_{tt}g_{ll}g_{\varphi\varphi}},\\\\
	\delta&=\frac{(g^{tt}g'_{tt}-g^{\varphi\varphi}g'_{\varphi\varphi})s}{2\sqrt{-g_{tt}g_{ll}g_{\varphi\varphi}}},\\\\
	\sigma&=-g^{\varphi\varphi}+\frac{g^{tt}(g'_{tt})^2s^2}{4g_{tt}g_{ll}g_{\varphi\varphi}},\\\\
	\gamma&=m^2\left(1-\frac{g'_{tt}g'_{\varphi\varphi}s^2}{4g_{tt}g_{ll}g_{\varphi\varphi}}\right)^2.
	\end{aligned}
	\end{equation}
	Equation~(\ref{3.25}) can be expressed as 
	\begin{equation}
	\label{3.27}
	(p_l)^2=\frac{\beta}{\alpha}(E-V_+)(E-V_-),
	\end{equation}
	where $V_\pm$ is a solution of $(p_r)^2=0$ given by  
	\begin{equation}
	\label{3.28}
	V_{\pm}=-\frac{\delta L}{\beta}\pm\sqrt{\frac{\delta^2 L^2}{\beta^2}+\frac{\gamma-\sigma L^2}{\beta}}.
	\end{equation}
	According to Eq.~(\ref{3.27}), the energy of the particle must satisfy the conditions
	\begin{equation}
	\label{3.29}
	\begin{array}{ccc}
	E\in (-\infty,V_-]&\text{or}&E\in[V_+,\infty),
	\end{array}
	\end{equation} 
	in order to have $(p^l)^2\geq0$. In the following, we shall focus on the case in which test particles have positive energy and therefore explore the effective potential given by $V_{\text{eff}}=V_+$.
	
	\subsection{Superluminal bound\label{SecIIB}}

    As mentioned, the dynamical four-momentum $p^\alpha$ and the kinematical four-velocity $u^\alpha$ of a spinning particle are not always parallel. As a consequence, although  $p_\alpha p^\alpha = -m^2$ is satisfied, the normalization $u_\alpha u^\alpha = -1$ does not hold. As the spinning particle moves closer to the center of symmetry, $u^\alpha$ increases, and eventually, for certain values of the spin $s$ and radius $l$, some components of the four-velocity may even diverge. Before this happens, the motion of the particle crosses the boundary between time-like and space-like trajectories thus becoming superluminal. It is well-known that space-like (or superluminal) motion does not have any physical meaning, and the transition to $u_\alpha u^\alpha > 0$ is not allowed for real particles. In this sense, one must impose a further constrain known as the superluminal bound, defined by the relation $u_\alpha u^\alpha = 0$. Hence, for the particle to move always in the time-like region, it is necessary to impose the following condition~\cite{Conde:2019juj, Toshmatov:2019bda} 
    \begin{equation}
    \label{3.30}
    \frac{u_\alpha u^\alpha}{(u^t)^2}=g_{tt}+g_{ll}(u^l)^2+g_{\phi\varphi}(u^\varphi)^2\leq 0,
    \end{equation}
    with the equality holding at the superluminal bound. To compute $u^l$ and $u^\varphi$, we use a method developed in Ref.~\cite{Hojman:2012me} (see App.~\ref{A1} for the full derivation). This method is based on application of the MPD equations (\ref{3.1}). From the second MPD equation, by using Tulczyjew-SSC, applying $D/d\lambda$ and solving it for $DS^{tl}/d\lambda$, $DS^{t\varphi}/d\lambda$ and $DS^{l\varphi}/d\lambda$ one can obtain the following system for the non-zero components of $S^{\alpha\beta}$:
    \begin{equation}
    \label{3.31}
    \begin{aligned}
    \frac{DS^{tl}}{d\lambda}&=p^tu^l-u^tp^l,\\
    \frac{DS^{t\varphi}}{d\lambda}&=p^tu^\varphi-u^tp^\varphi,\\
    \frac{DS^{l\varphi}}{d\lambda}&=p^lu^\varphi-u^lp^\varphi.
    \end{aligned}
    \end{equation}
    Following to the gauge choices and invariant relations in Ref.~\cite{Hojman:2012me}, we set $\lambda = t$ and can express the above system of equations in terms of $S^{\varphi l}$ only. This is a consequence of the MPD equations that imply
    \begin{equation}
    \label{3.44}
    \begin{aligned}
    u^l&=\frac{\hat{\mathcal{C}}}{\hat{\mathcal{B}}}\frac{p_l}{p_t},\\
    u^\varphi &=\frac{\hat{\mathcal{A}}}{\hat{\mathcal{B}}}\frac{p_\varphi}{p_t},
    \end{aligned}
    \end{equation}
    with
    \begin{equation}
    \label{3.45}
    \begin{aligned}
    \hat{\mathcal{A}}&=g^{\varphi\varphi}+R_{tllt}\left(\frac{S^{\varphi l}}{p_t}\right)^2,\\
    \hat{\mathcal{B}}&=g^{tt}+R_{\varphi ll\varphi}\left(\frac{S^{\varphi l}}{p_t}\right)^2,\\
    \hat{\mathcal{C}}&=g^{ll}+R_{\varphi tt\varphi}\left(\frac{S^{\varphi l}}{p_t}\right)^2.
    \end{aligned}
    \end{equation}
    
    Now, inserting Eq.~(\ref{3.44}) into the superluminal bound condition \eqref{3.30} we have
    \begin{equation}
    \label{3.47}
    g_{tt}(\hat{\mathcal{B}})^2(p_t)^2+g_{ll}(\hat{\mathcal{C}})^2(p_l)^2+g_{\varphi\varphi}(\tilde{A})^2(p_\varphi)^2\leq 0,
    \end{equation}
    and from the conservation of the four-momentum $p_\alpha p^\alpha=-m^2$, the superluminal bound condition reduces to
    \begin{equation}
    \label{3.48}
    \mathcal{F}=\left(\frac{p_t}{m}\right)^2\mathcal{X}+\left(\frac{p_\varphi}{m}\right)^2\mathcal{Y}-\mathcal{Z}\leq 0,
    \end{equation}
    with 
    \begin{equation}
    \label{3.49}
    \begin{aligned}
    \mathcal{X}&=g_{tt}(\hat{\mathcal{B}})^2-\frac{(\hat{\mathcal{C}})^2}{g_{tt}},\\
    \mathcal{Y}&=g_{\varphi\varphi}(\hat{\mathcal{A}})^2-\frac{(\hat{\mathcal{C})^2}}{g_{\varphi\varphi}},\\
    \mathcal{Z}&=g_{ll}(\hat{\mathcal{C}})^2.
    \end{aligned}
    \end{equation}
 
    \subsection{Other bounds on the spin parameter \label{SecIIIC}}
    When dealing with realistic atrophysical scenarios, such as compact stars and stellar mass black holes orbiting supermassive black holes, we must notice that the approximations considered here have limits. The mass and size of the spinning test particle must be negligible with respect to the central object's mass and must not affect the geometry. Furthermore, it is crucial to remark that MPD equations only take into account effects generated by the mass monopole and spin dipole, the well-known pole-dipole approximation. In this sense, these equations do not take into account the mass quadrupole effect related to the tidal coupling. Hence, one can not use them to model more extended objects~\cite{Semerak:1999qc, Hartl:2002ig}. Additionally, since the MPD equations are conservative, the effects of gravitational radiation are ignored.
    
    In this sense, following Refs.~\cite{Semerak:1999qc, Hartl:2002ig}, it is important to point out that we measure the spin parameter $s$ in terms of $mb_0$, not $m^2$. Therefore, the system considered in this work is a spinning compact body of mass $m$ orbiting a super massive wormhole of mass $b_0$. In the case of compact objects, such as black holes, neutron stars, or white dwarfs, physically realistic values of the spin must satisfy $s<<1$. If the compact object is a maximally spinning black hole of mass $m$, with spin angular momentum $m^2$ (and $m<<b_0$), for example, the spin parameter $s$ is given by the condition~\cite{Hartl:2002ig} 
    \begin{equation}
    \label{3.50}
    s=\frac{S}{m b_0}\leq\frac{m^2}{mb_0}=\frac{m}{b_0}<<1.
    \end{equation}
    On the other hand, if the compact object is a neutron star, one can take a maximum value (approximately) of $0.6m^2$~\cite{Cook:1993qr}. From which one gets $s\lesssim 0.6m/b_0$.

    When the compact object is a white draft, the bound on the spin is more difficult to compute. First, one needs to consider the maximum value of $s$ before the star begins to break up. This value is given by $s_{max}=I\Omega_{max}$, where $\Omega_{max}$ is the maximum angular velocity and $I$ is the moment of inertia. If we we write $I=\alpha m R^2 $ and $\Omega_{max}=\beta\sqrt{m/R^3}$ for some constants $\alpha=0.2044$ and $\beta\lesssim1$ ($\beta=0.5366$), we have that~\cite{Hartl:2002ig} 
    \begin{equation}
    \label{3.51}
    s_{max}=0.110\sqrt{m^3R}.
    \end{equation}     
    Nevertheless, the limit in Eq.~(\ref{3.51}), depends on the mass-radius ratio between the objects in question. Therefore, an analytical expression may be necessary to compute the mass-radius ratio. One example is given by the relation (for non rotating white dwarfs)~\cite{Hartl:2002ig} 
    \begin{equation}
    \label{3.52}
    \frac{R}{R_\odot}=0.1125\left(\frac{m}{m_{max}}\right)^{-\frac{1}{3}}f^\frac{1}{2}(m),
    \end{equation} 
    with
    \begin{equation}
    \label{3.53}
    f(m)=1-\left(\frac{m}{m_{max}}\right)^\frac{4}{3},
    \end{equation}
    and $m_{max}=1.454M_\odot$. From which, following the results of Ref.~\cite{Lai:1993pa}, it is possible to obtain the limit
    \begin{equation}
    \label{3.54}
    s\leq s_{max}=\frac{S_{max}}{mb_0}=9\times 10^{-6}<<1. 
    \end{equation} 


	\section{Spinning particles around the Morris-Thorne wormhole\label{SecIII}} 
	
    In this section, we use the results from Sec.~\ref{SecII} to compute the effective potential $V_{\text{eff}}$ for a spinning test particle moving around a wormhole. As we have shown, the effective potential depends on $\alpha$, $\beta$, $\delta$, $\sigma$, $\gamma$, which are given by Eqs.~(\ref{3.26}). Note that these functions depend on the redshift function $\Phi$. 
    As mentioned before, we focus on particles with positive energy taking $V_{\text{eff}}=V_+$ in Eq.~(\ref{3.28}).
    
    In general it is useful to rescale all the quantities in such a way that they are adimensional.
    Hence, we define
    \begin{equation}
    \label{4.1}
    \begin{array}{ccc}
    l\rightarrow\frac{l}{b_0},&s\rightarrow\frac{s}{b_0}=\frac{S}{m b_0},&\mathcal{L}\rightarrow\frac{\mathcal{L}}{b_0}=\frac{L}{m b_0}.
    \end{array}
    \end{equation}
    Since the throat of the wormhole can be interpreted as the gravitational mass of the source, we can rescale the radial distances per unit mass or equivalently set $b_0=M=1$. Therefore, from 
    Eqs.~(\ref{3.26}), and (\ref{3.28}) we obtain
	\begin{figure*}[t]
		\begin{center}
		$   
		\begin{array}{ccc}
			\includegraphics[scale=0.3]{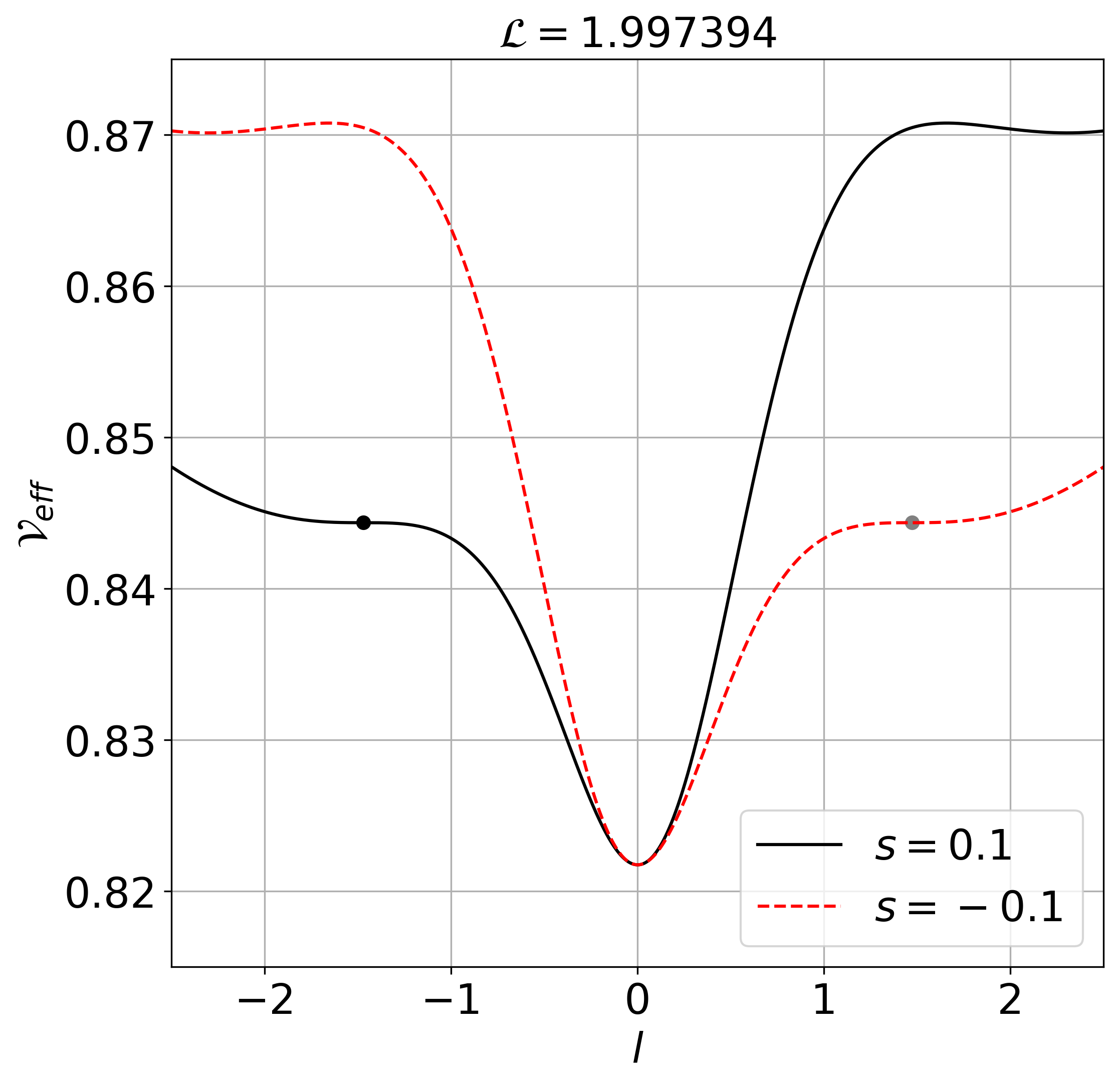} &
			\includegraphics[scale=0.3]{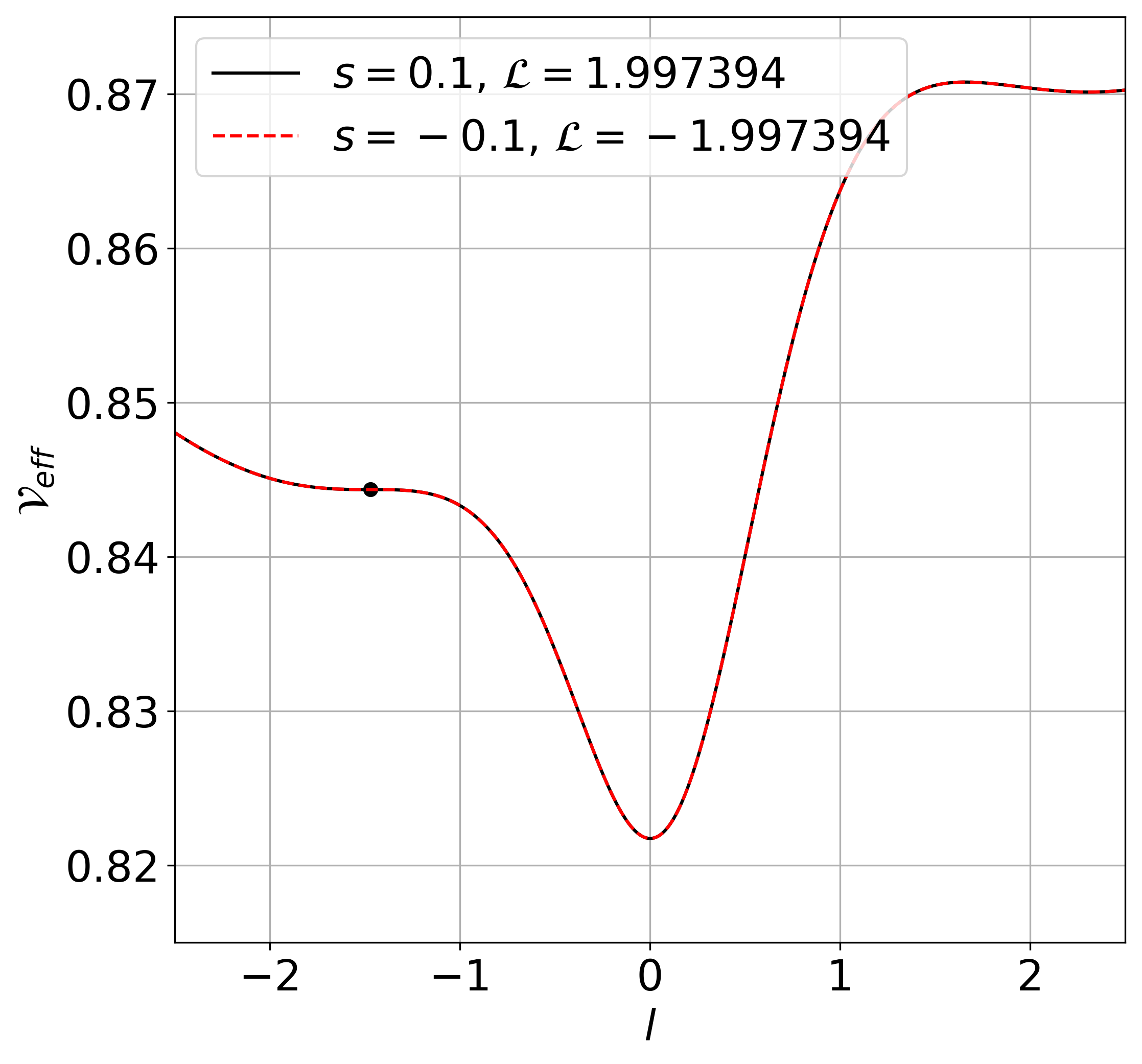}&
			\includegraphics[scale=0.3]{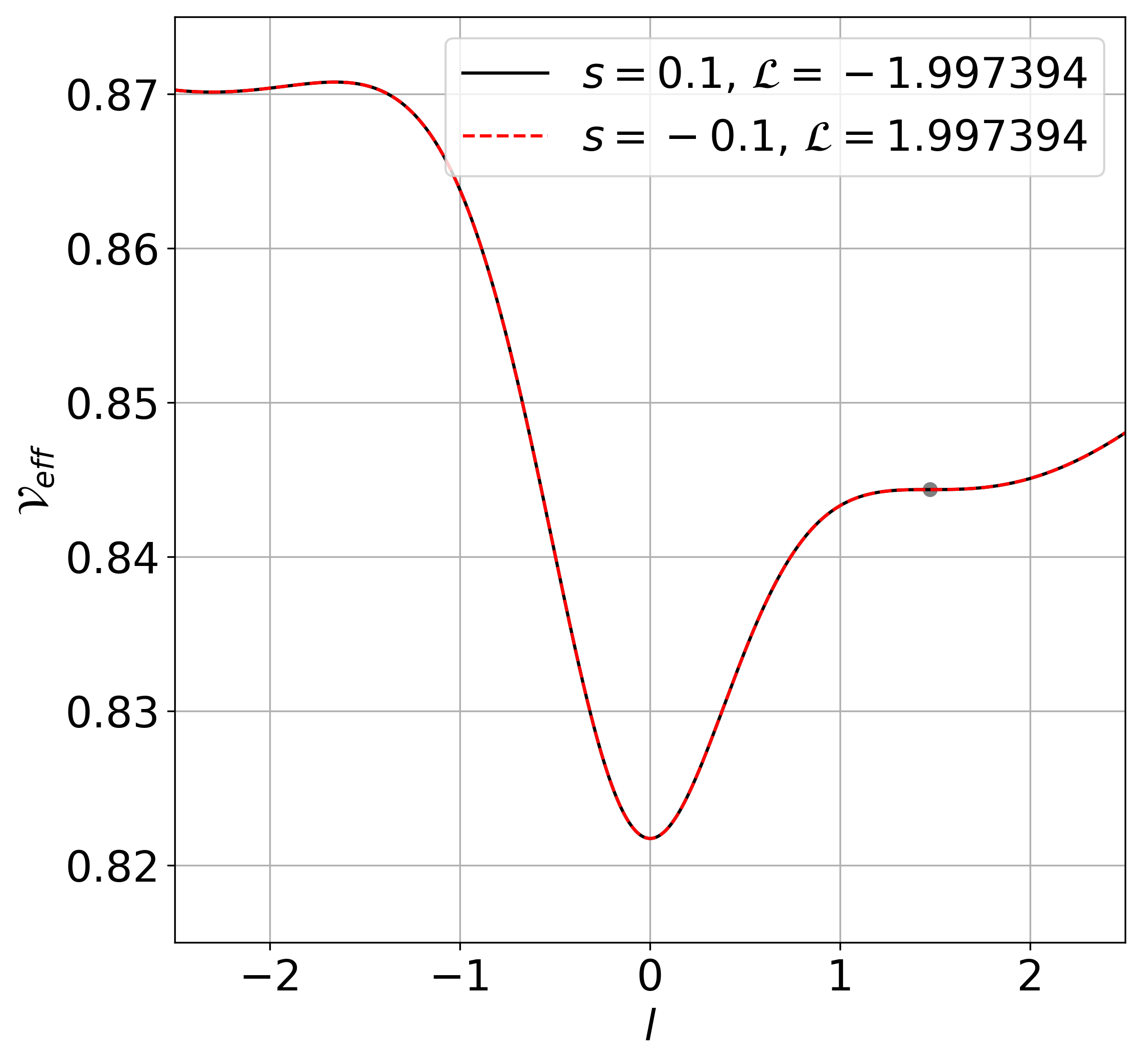} \\
			\end{array}
		$
		\end{center}
		\caption{The effective potential $\mathcal{V}_{\text{eff}}$ as a function of the radial coordinate $l$ with the throat located at $l=0.$ Left panel: Plots of $\mathcal{V}_{\text{eff}}$ in ``plus''(red) and ``minus'' (black) configurations. Central panel: two ``plus'' configurations. Right panel: two ``minus'' configurations. The dots show the location of the ISCO in the lower (black) and upper (gray) universes. In the plots we assume $b_0=M=1$.\label{fig3}}
	\end{figure*}
    \begin{equation}
    \label{4.2}
    \begin{aligned}
    \mathcal{V}_{\text{eff}}(l,s,\mathcal{L})&=-\frac{\delta\mathcal{L}}{\beta}+\sqrt{\left(\frac{\delta\mathcal{L}}{\beta}\right)^2+\frac{\gamma-\mathcal{L}^2\sigma}{\beta^2}},
    \end{aligned}
    \end{equation}
    where we have also defined the adimensional effective potential per unit mass of test particle by setting $\mathcal{V}_{\text{eff}}\rightarrow V_{\text{eff}}/m$. With the above adjustments we get
    \begin{equation}
    \label{4.3}
        \begin{aligned}
        \beta&=\frac{e^{\frac{2}{\sqrt{1+l^2}}}[1+l(l-s)][1+l(l+s)]}{(1+l^2)^2},\\
        \delta&\rightarrow b_0\delta=\frac{ls(1-\sqrt{1+l^2})}{(1+l^2)^{\frac{3}{2}}\sqrt{e^{-\frac{2}{\sqrt{1+l^2}}}(1+l^2)}},\\
        \sigma&\rightarrow b^2_0\sigma=\frac{l^2s^2}{(1+l^2)^4}-\frac{1}{1+l^2},\\
        \gamma&\rightarrow\frac{\gamma}{m^2}=\left(1-\frac{l^2s^2}{(1+l^2)^{\frac{5}{2}}}\right)^2.
        \end{aligned}
    \end{equation}
    From Eqs.~(\ref{4.2}) and (\ref{4.3}), it is possible to see the symmetries in $\mathcal{V}_{\text{eff}}(l)$ depending on the signs of $s$, and $\mathcal{L}$. 
    We may call $\mathcal{V}^P_{\text{eff}}$ the ``plus'' configuration with positive signs for $s$ and $\mathcal{L}$. Then we have
    \begin{equation}
    \label{4.4a}
        \begin{aligned}
            \mathcal{V}^P_{\text{eff}}(l,s,\mathcal{L})&=\mathcal{V}^P_{\text{eff}}(l,-s,-\mathcal{L}).
        \end{aligned}   
    \end{equation}
    Also we may call $\mathcal{V}^M_{\text{eff}}$ the ``minus'' configuration with the opposite sign in the first term of Eq.~\eqref{4.2}. This is obtained when either $\mathcal{L}$ or $\delta$ or $\beta$ changes sign. Then we have 
    \begin{equation}
    \label{4.4}
        \begin{aligned}
            \mathcal{V}^M_{\text{eff}}(l,s,-\mathcal{L})&=\mathcal{V}^M_{\text{eff}}(l,-s,\mathcal{L}).
        \end{aligned}   
    \end{equation}
    \begin{table}
    \begin{center}
    \begin{tabular}{||c c c c c c c c||} 
    \hline
    $l$ & $s$ & $\mathcal{L}$& Configuration & $l$ & $s$ & $\mathcal{L}$& Configuration \\ [0.5ex] 
    \hline\hline
    + & + & + & Plus & - & - & - & Plus\\ 
    \hline
    + & + & - & Minus & - & - & + & Minus\\
    \hline
    + & - & + & Minus & - & + & - & Minus\\
    \hline
    + & - & - & Plus  & - & + & + & Plus\\
    \hline
    \end{tabular}
    \caption{Symmetries and configurations for $\mathcal{V}_{\text{eff}}$ depending on the signs of $l$, $s$, and $\mathcal{L}$.\label{tab1}}
    \end{center}
    \end{table}
    In Table.~\ref{tab1} we show in detail the symmetries for $\mathcal{V}_{\text{eff}}$ in the upper ($l>0$) and lower ($l<0$) universes. Moreover, the behavior of each configuration is shown in Fig.~\ref{fig3}. In the left panel, we have plot together $\mathcal{V}^M_{\text{eff}}(l,-0.1,1.997394)$ (red) and $\mathcal{V}^P_{\text{eff}}(l,0.1,1.997394)$ (black), as functions of $l$. Note that the change in the configuration from ``plus'' to ``minus'' also changes the location of the innermost stable circular orbit (ISCO), shown in the figure with black and gray dots. In the central panel, we plot the ``plus'' configuration given by $\mathcal{V}^P_{\text{eff}}(l,0.1,1.997394)=\mathcal{V}^P_{\text{eff}}(l,-0.1,-1.997394)$. Finally, in the right panel, we plot the ``minus'' configuration given by
    $\mathcal{V}^M_{\text{eff}}(l,-0.1,1.997394)=\mathcal{V}^M_{\text{eff}}(l,0.1,-1.997394)$.
		\begin{figure}[h!]
		\begin{center}
			\includegraphics[scale=0.37]{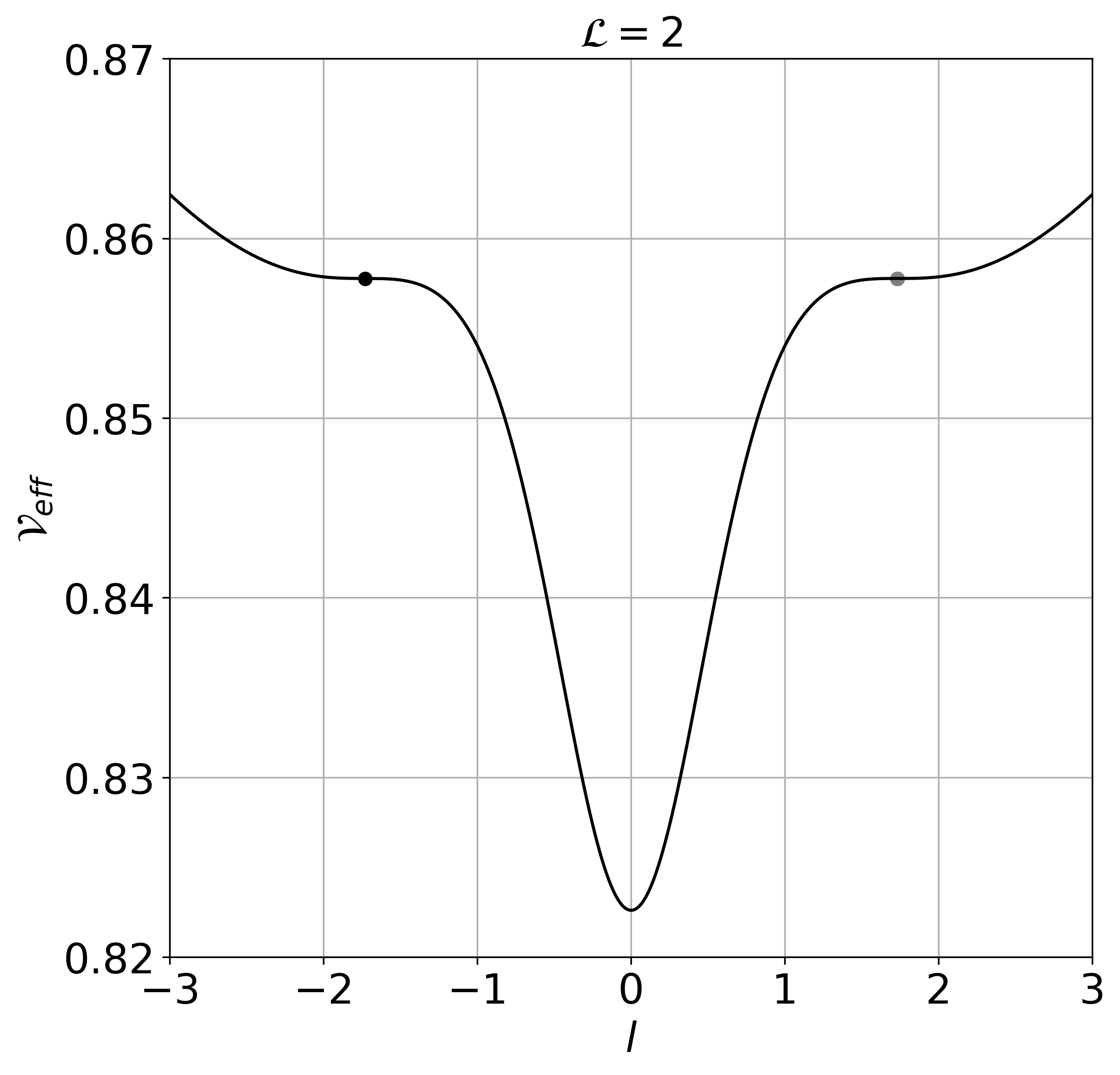}
			\caption{The effective potential for a non-spinning test particle ($\tilde{s}=0$) is symmetric about the throat, hence the ISCO for non-spinning particles is at the same distance from the throat in the upper and lower universes. In the plots we assume $b_0=M=1$.\label{fig4}}
		\end{center}
	\end{figure}
	
	\begin{figure*}[t]
		\begin{center}$
			\begin{array}{ccc}
			\includegraphics[scale=0.3]{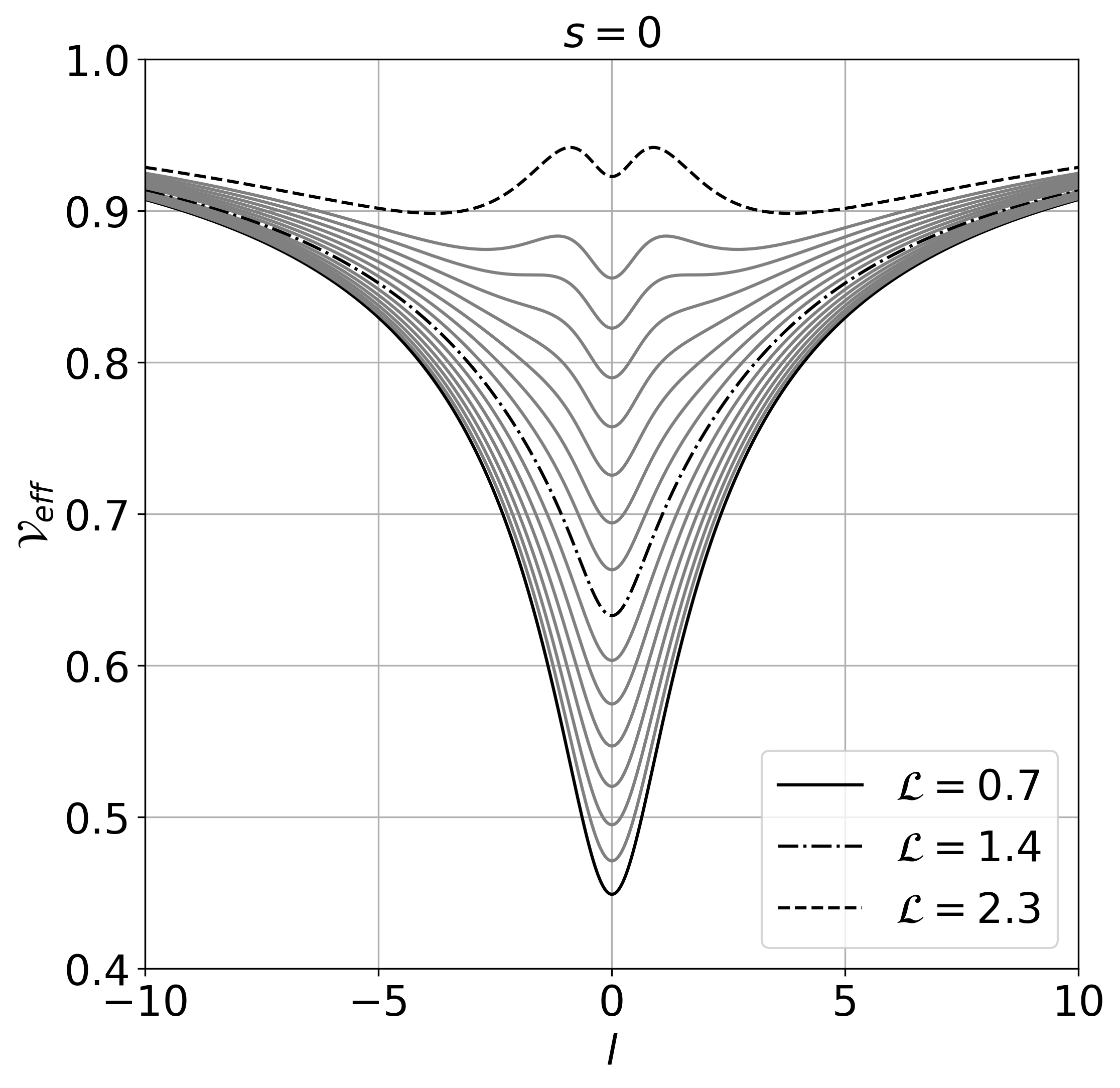} &
			\includegraphics[scale=0.3]{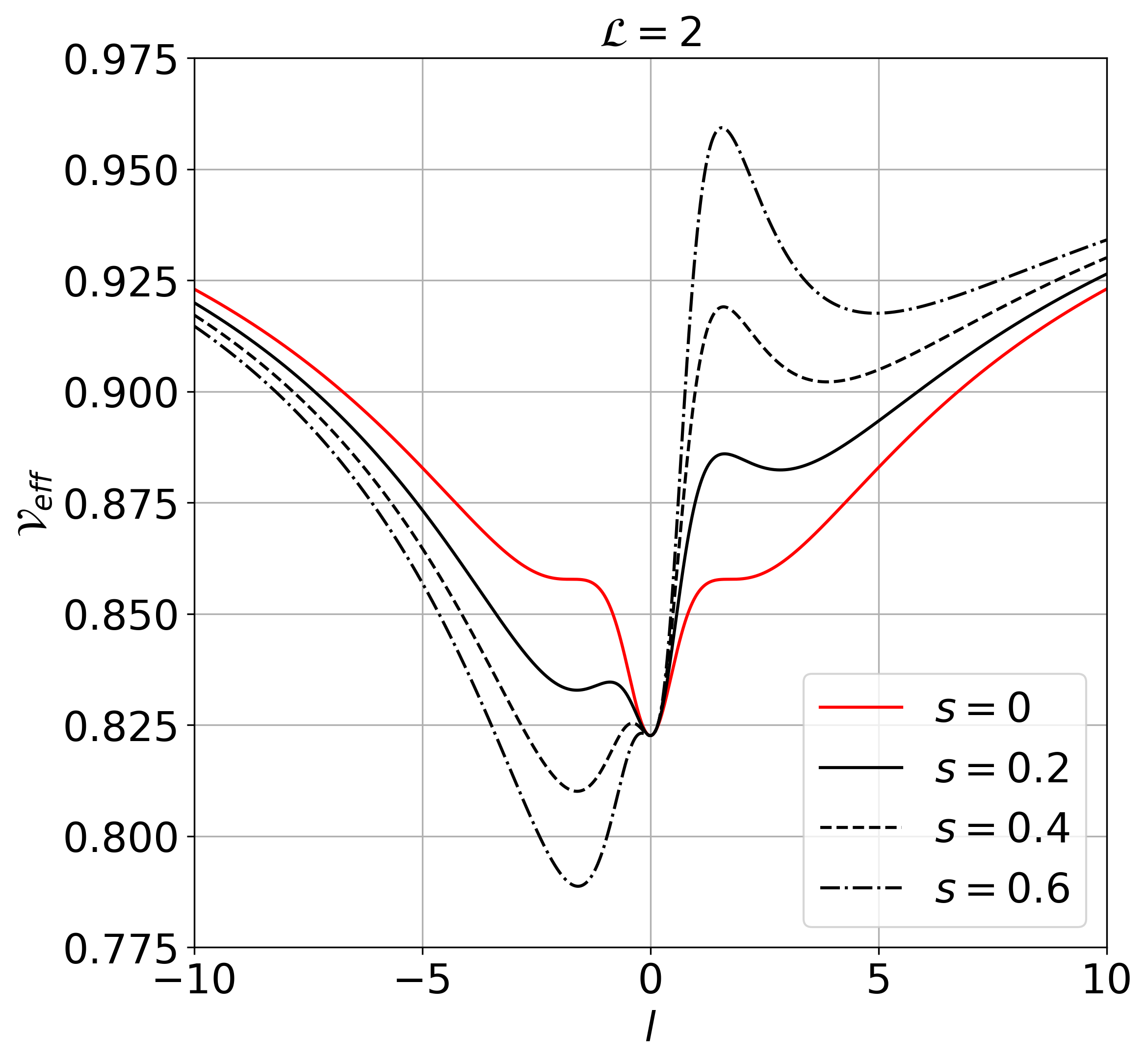}&
			\includegraphics[scale=0.3]{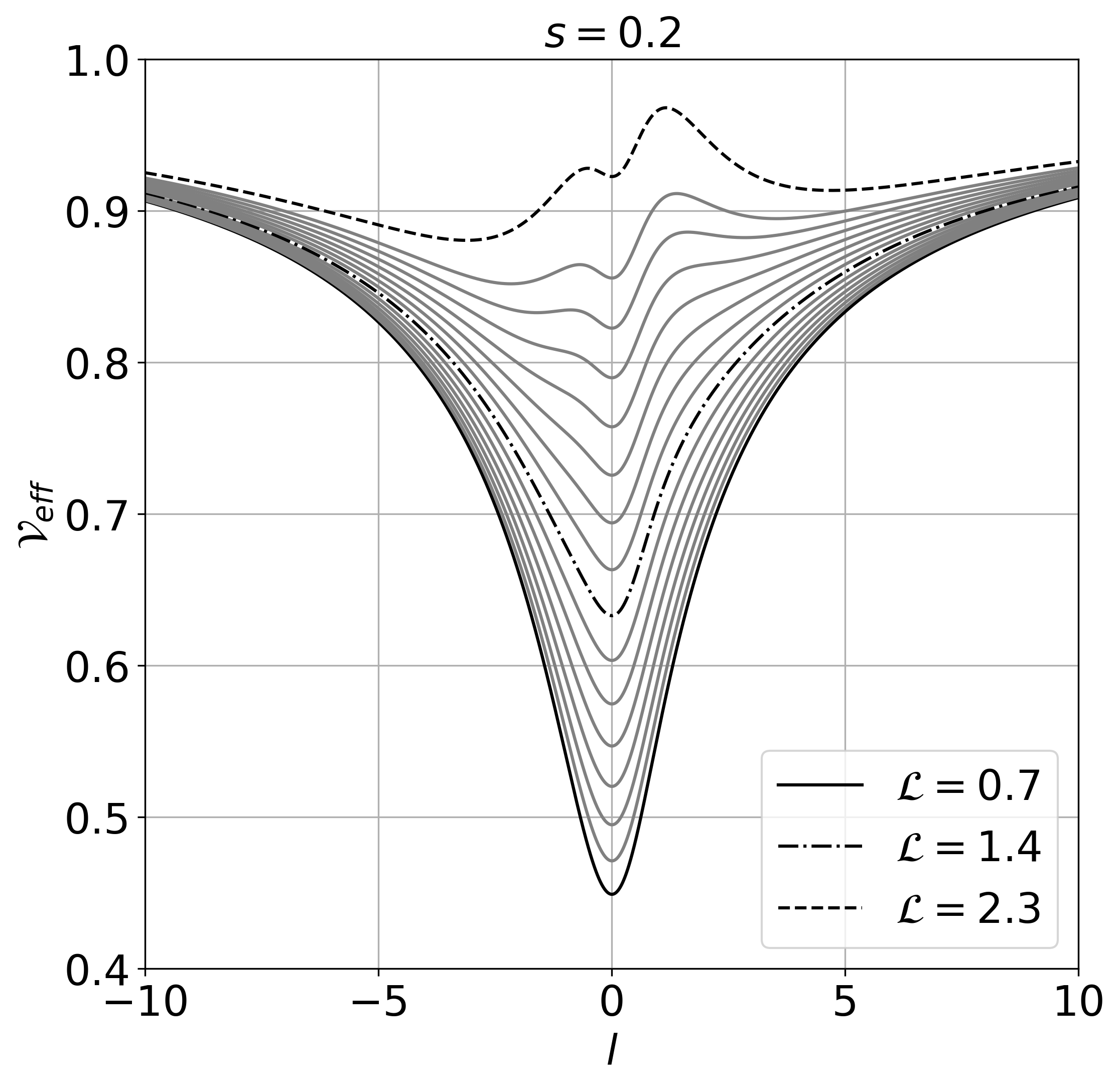} \\
			\end{array}$
		\end{center}
		\caption{Plots of $\mathcal{V}_{\text{eff}}$ as a function of $l$ in different situations. Left panel: for a non-rotating particle with different values of $\mathcal{L}$. Central panel: constant angular momentum $\mathcal{L}=2$ and different values of $s$. Right panel: constant spin $s=0.2$ with different values of $\mathcal{L}$. In the plots we consider $b_0=M=1$. Furthermore, in the the left and right panels, we vary $\mathcal{L}$ from $0.7$ to $2.3$ by a step of $0.1$.\label{fig5}}
	\end{figure*}
	
    One interesting feature one sees from Fig.~\ref{fig3} is that the profile of the effective potential is not symmetric about the throat $l=0$ due to the spin of the particle, in contrast to the motion of non-spinning particles ($s=0$) as shown in Fig.~\ref{fig4}. The direction of the spin of the test particle changes the shape of the effective potential removing its symmetry between upper and lower universe. In the ``plus'' configuration, for example, the value of $\mathcal{V}^P_{\text{eff}}$ in the upper universe is larger than that in the lower universe. On the other hand, in the ``minus'' configuration, the value of the effective potential in the lower universe is larger than that in the upper universe, as can be seen from Fig.~\ref{fig3}. For spinning particles the symmetry of $\mathcal{V}_{\text{eff}}$ depends on which universe (i.e. sign of $l$) and direction of the spin (i.e. sign of $s$), so that particles with positive spin in the upper universe have the same motion as particles with negative spin in the lower universe. 
    
	\begin{figure*}[h!]
		\begin{center}$
			\begin{array}{ccc}
			\includegraphics[scale=0.3]{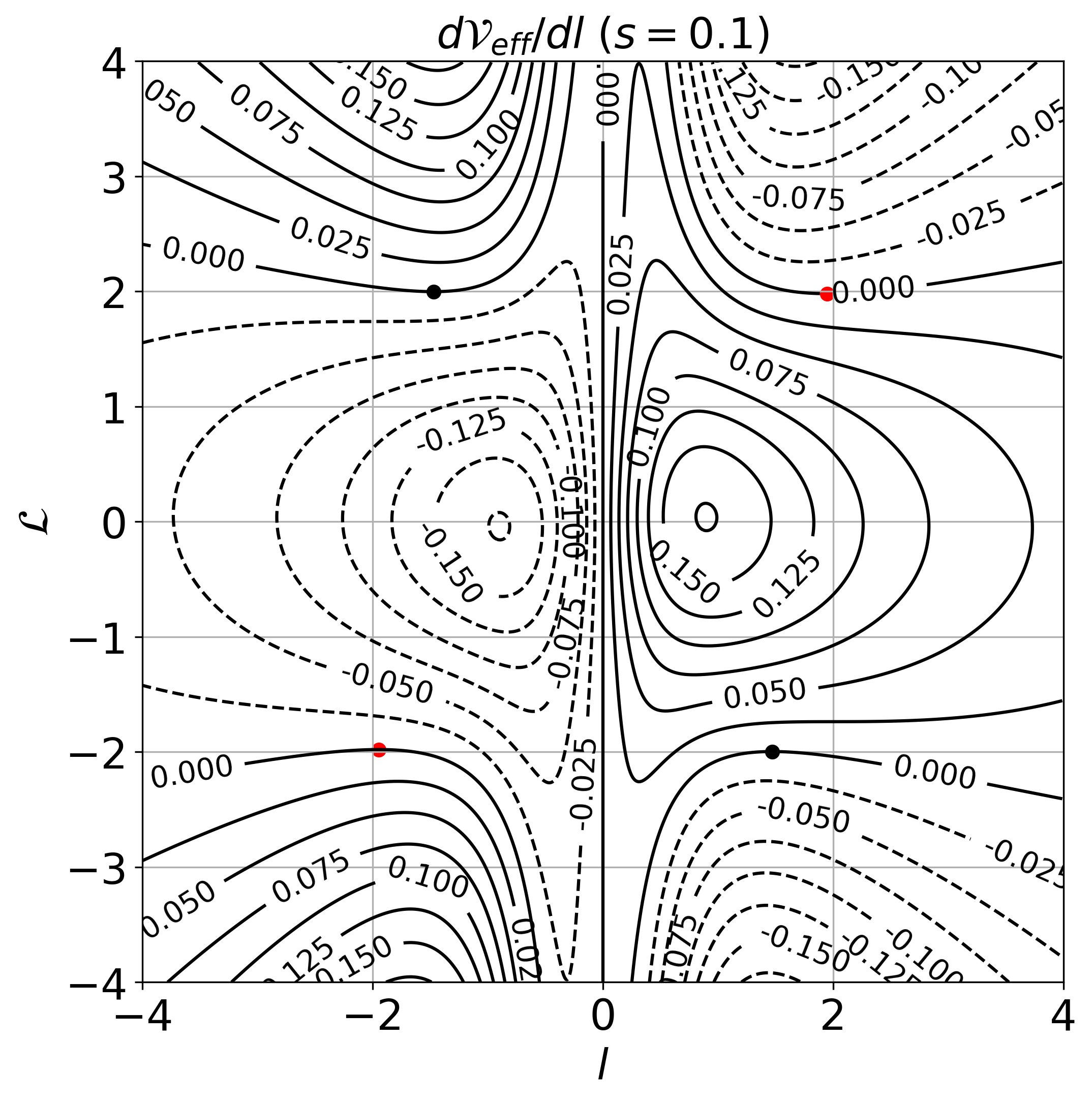} &
			\includegraphics[scale=0.3]{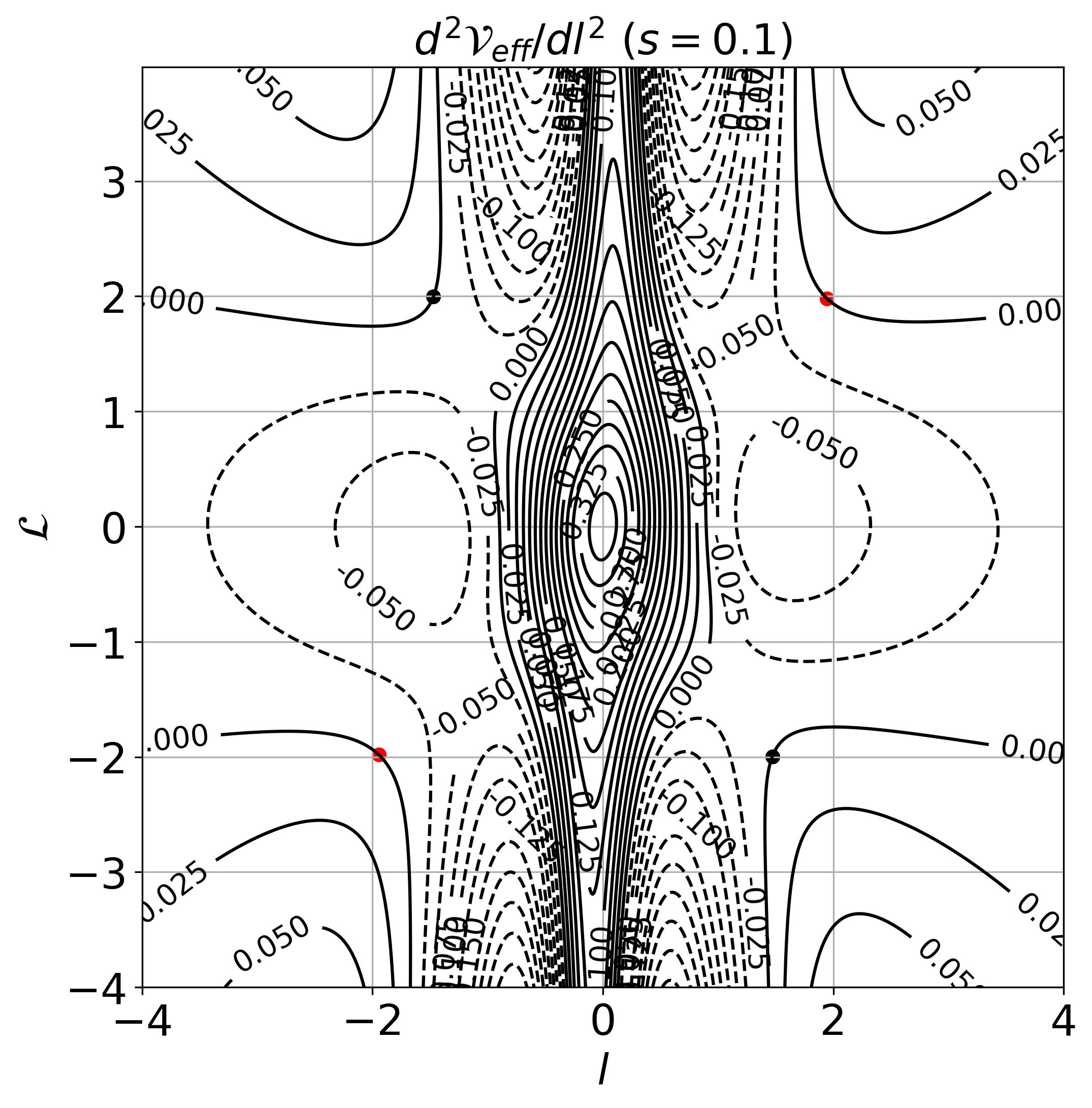}&
			\includegraphics[scale=0.3]{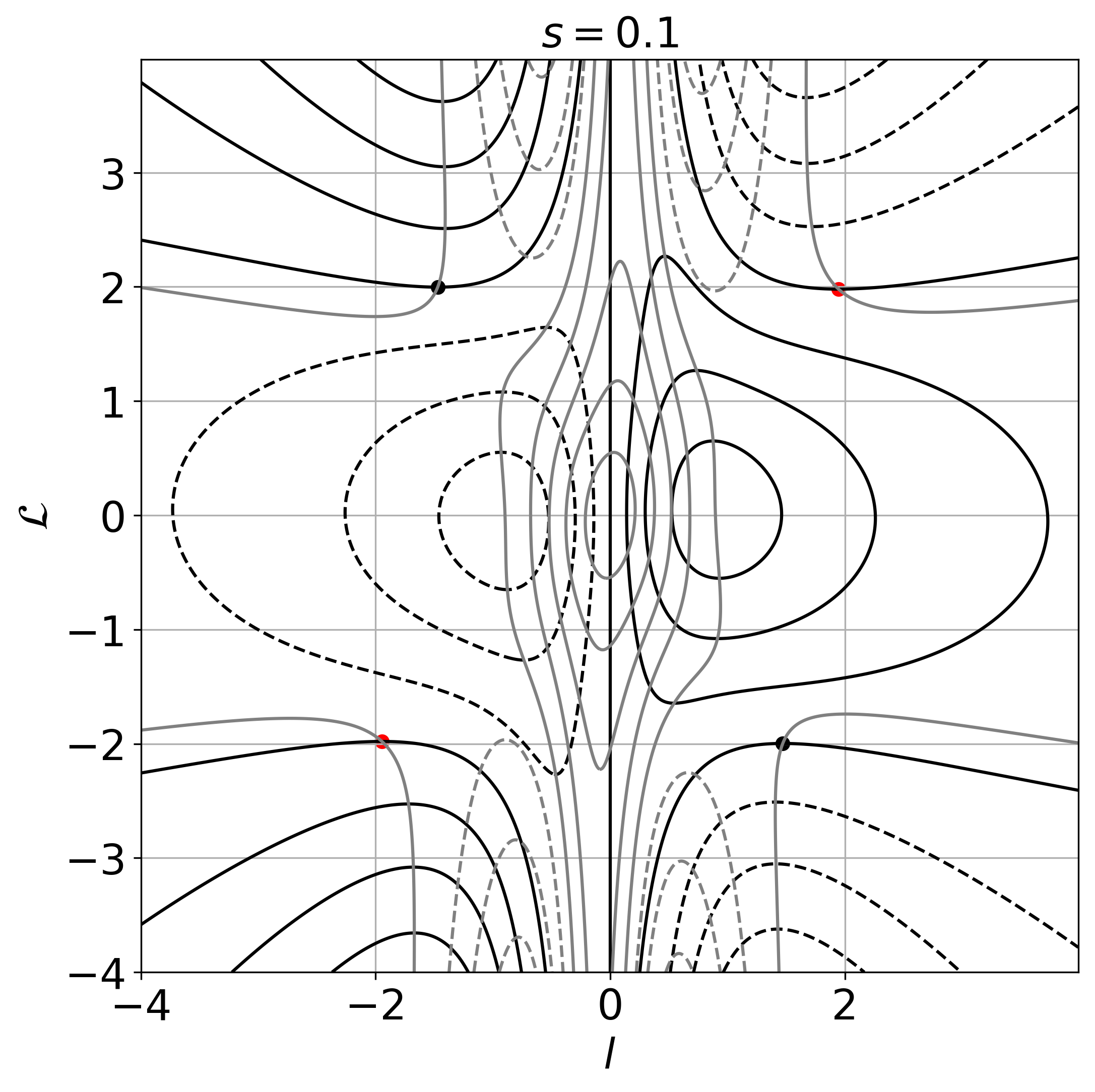} \\
			\includegraphics[scale=0.3]{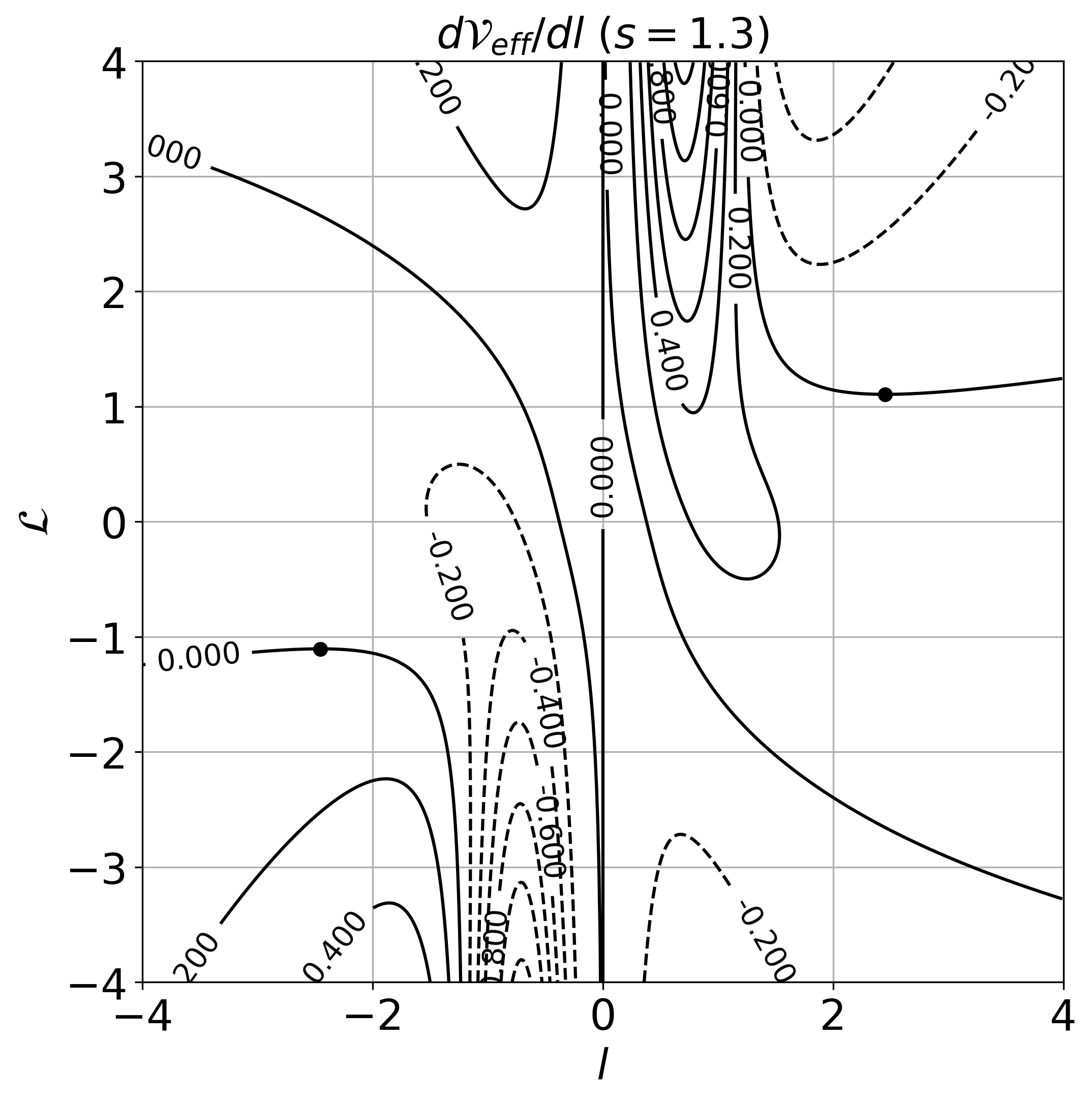} &
			\includegraphics[scale=0.3]{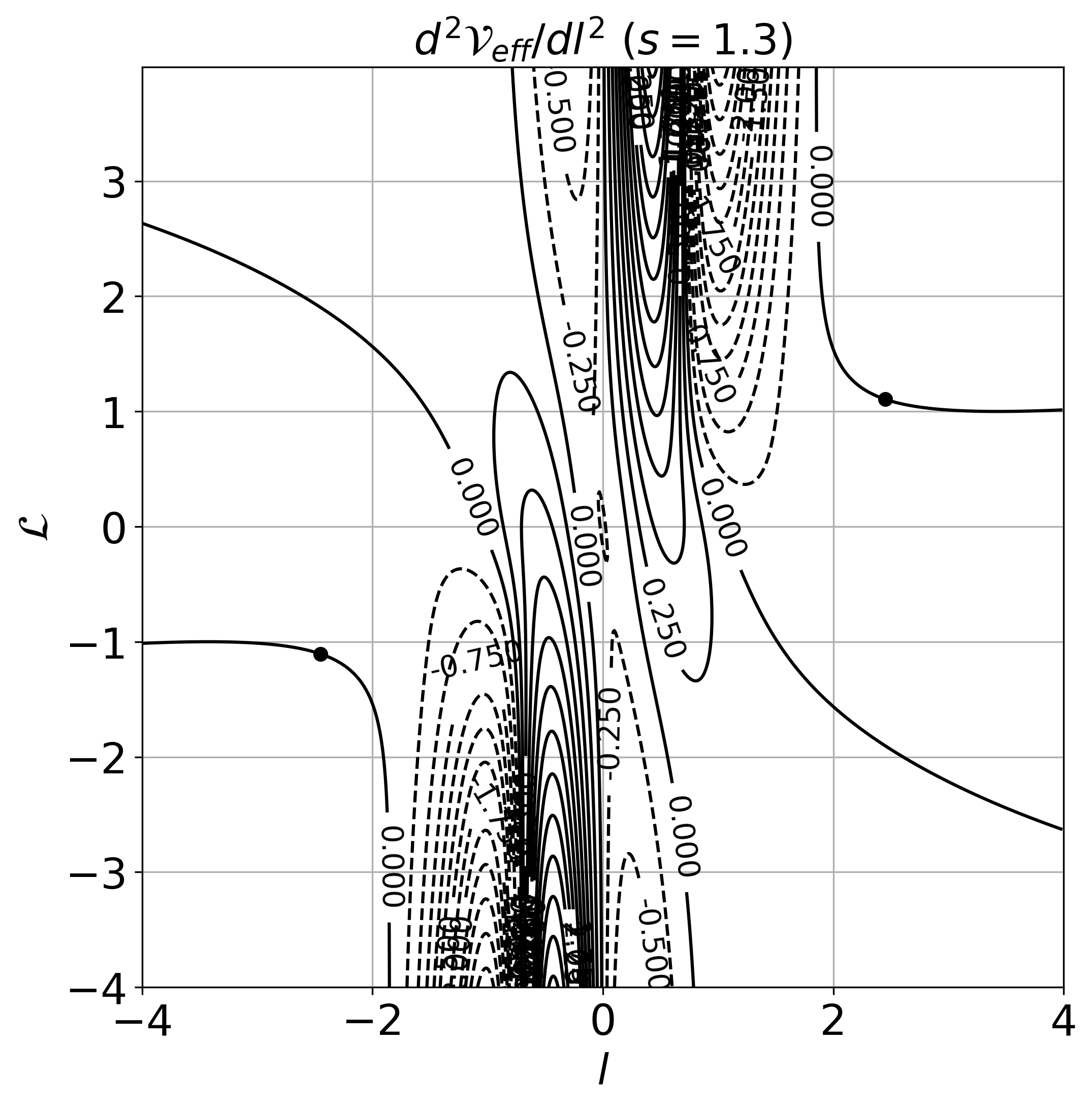}&
			\includegraphics[scale=0.3]{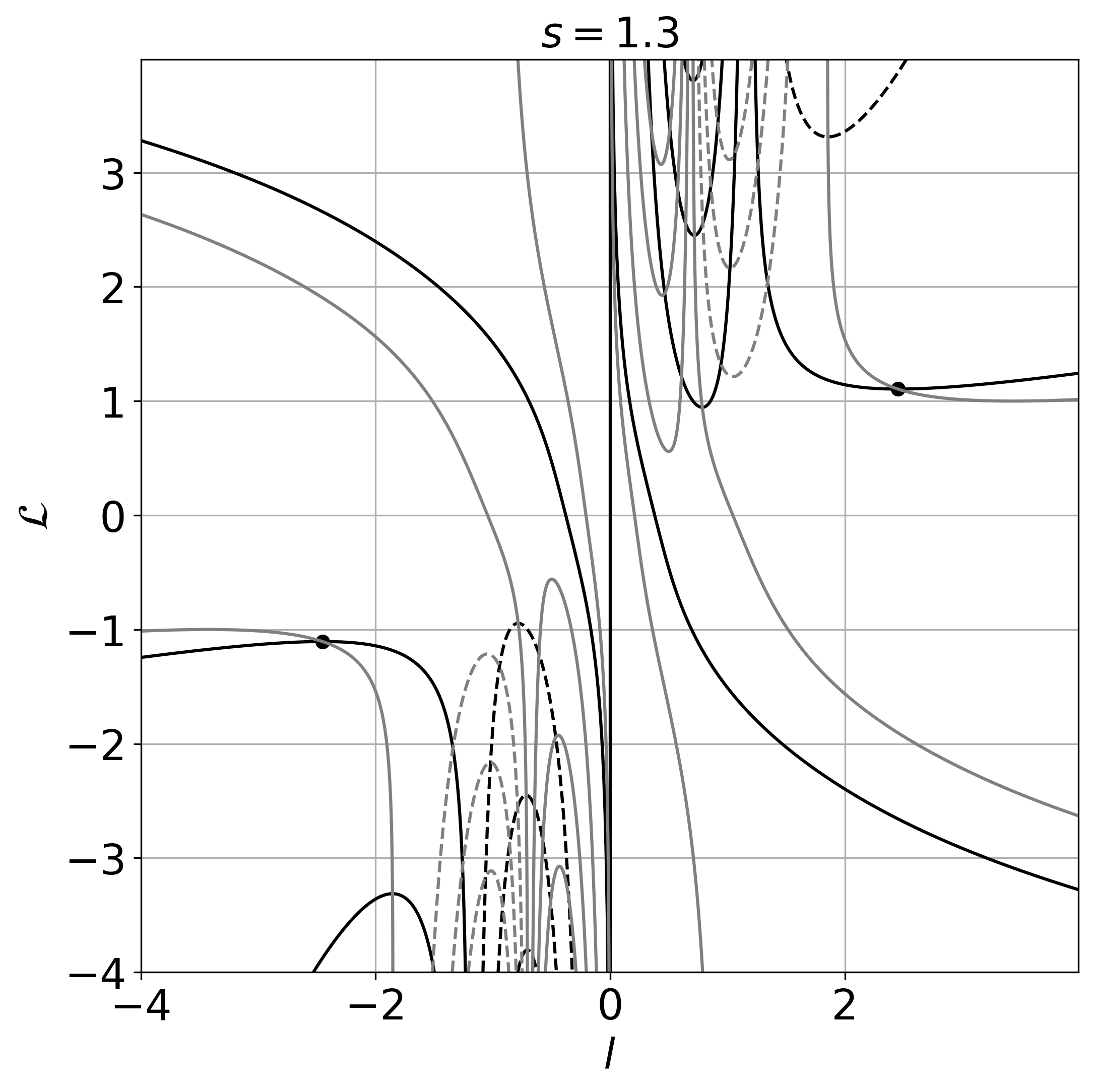} 
			\end{array}$
		\end{center}
		\caption{Contour plots for the first and second derivatives of $\mathcal{V}_{\text{eff}}(l,\mathcal{L})$ and location of the ISCO shown using dots with black a red colors. In the first row we have $s=0.1$, in the second row $s=1.3$. Left panel: Contour plot of $d\mathcal{V}_{\text{eff}}/dl={\rm const.}$. The ISCO is located on the curve $d\mathcal{V}_{\text{eff}}/dl=0$. Middle panel: Contour plot of $d^2\mathcal{V}_{\text{eff}}/dl^2={\rm const.}$. The ISCO is located on the curve $d^2\mathcal{V}_{\text{eff}}/dl^2=0$. Right panel: Superposition of $d\mathcal{V}_{\text{eff}}/dl$ and $d^2\mathcal{V}_{\text{eff}}/dl^2$. The ISCO (when the first and second derivatives vanish) is given by the point of intersection of the two curves $d\mathcal{V}_{\text{eff}}/dl=d^2\mathcal{V}_{\text{eff}}/dl^2=0$. In the plots we assume $b_0=M=1$.\label{fig6}}
	\end{figure*}

    In Fig.~\ref{fig5}, the behavior of $\mathcal{V}_{\text{eff}}$ as function of $l$ in different situations is plotted. In the left panel, we show the effective potential for a non-rotating particle with different values of the angular momentum $\mathcal{L}$, varying from $0.7$ to $2.3$ and increasing with a step of $0.1$. In the figure, it is possible to see the symmetry about the throat in the shape of $\mathcal{V}_{\text{eff}}$, which has the same behaviour in both the lower and upper universes. As the angular momentum increases, the effective potential increases. For large values of $\mathcal{L}$, such as $\mathcal{L}=2.1$, $2.2$ and $2.3$, the behavior of $\mathcal{V}_{\text{eff}}$ shows two maxima (unstable circular orbits) and three minima, one of which is located at the throat. The behavior in the upper (lower) universe outside the throat resembles the usual behavior in the Schwarzschild geometry. Also, when the angular momentum $\mathcal{L}$ decreases, the two peaks vanish, and the shape of $\mathcal{V}_{\text{eff}}$ changes in such a way that only one minimum value appears at the throat. One can see similar behavior in the right panel of Fig.~\ref{fig5}, where the effective potential of a spinning particle ($s=0.2$) for different values of $\mathcal{L}$ is plotted. However here we see that the presence of the spin removes the symmetry of $\mathcal{V}_{\text{eff}}$ about the throat and the effective potential does not have the same behavior in both universes. This is clearly highlighted in the central panel of Fig.~\ref{fig5} where $\mathcal{V}_{\text{eff}}$ is plotted for various values of $s$, while keeping the angular momentum ($\mathcal{L}=2$) constant. As above mentioned, the profile of $\mathcal{V}_{\text{eff}}$ is symmetric only when the spin of the particle is zero (red color curve in the central panel of Fig.~\ref{fig5}). Finally, it is important to point out that  $\mathcal{V}_{\text{eff}}$ tends to $1$ when $l\rightarrow\pm\infty$.
    
    \section{Innermost stable circular orbits\label{Section IV}}
    
    Now, we focus our attention on circular orbits of spinning particles in the space-time of a wormhole given by Eq.~(\ref{3.11a}). Circular motion occurs when the radius is constant and the radial acceleration of the particle vanishes. Mathematically, this means that the radial velocity of the test particle vanishes $dl/d\lambda=0$, which implies that $\mathcal{E}=\mathcal{V}_{\text{eff}}(l)$ (see Eq.~(\ref{3.27})), where we have defined the energy of the test particle per unit mass as $\mathcal{E}=E/m$. The radial acceleration of the particle also vanishes, i.e. $d^2l/d\lambda^2=0$, from which $d\mathcal{V}_\text{eff}/dl=0$. Nevertheless, this condition does not guarantee that circular orbits are stable. The stability of a circular orbit can be evaluated from the second derivative of the effective potential with respect to the radial coordinate, which must be positive, namely
	\begin{equation}
	\label{4.1}
	\frac{d^2\mathcal{V}_\text{eff}}{dl^2}\geq 0.
	\end{equation} 
	When $d^2\mathcal{V}_\text{eff}/dl^2=0$, one can obtain the marginally stable circular orbit, corresponding to the smallest allowed value for stable circular orbits, also known as the innermost stable circular orbit or ISCO.
	
	Using the conditions $\mathcal{E}=\mathcal{V}_{\text{eff}}$ and $d\mathcal{V}_\text{eff}/dl=0$ we obtain the values of $\mathcal{E}$ and $\mathcal{L}$ in terms of the circular orbit radius $l$ and then from $d^2\mathcal{V}_\text{eff}/dl^2=0$ the value of the radius of the ISCO for a spinning test particle. Since the process involves the solution of a non-linear system of equations for $l$ and $\mathcal{L}$, we must in general solve it numerically. Also, to ensure that motion of the spinning particle is physically valid, we check that the superluminal bound is satisfied using Eq.~(\ref{3.48}).
	
	In Fig.~\ref{fig6}, we show the contour plots for the first and second derivatives of the effective potential for two different values of the spin $s=0.1$ and $1.3$. When we consider $s=0.1$ and the ``plus'' configuration (both $\mathcal{L}$ and $s$ positives), the contour plot shows two possible values for the ISCO: one located in the upper universe ($l_\text{ISCO}=1.946354727$) and the other is in the lower universe ($l_{\text{ISCO}}=-1.470809497$). It is important to point out that each value of the ISCO corresponds to a different value in the angular momentum. Hence, in the upper universe, a particle with spin $s=0.1$ moves in a stable circular orbit at the ISCO radius $l=1.946354727$ with angular momentum $\mathcal{L}=1.979967293191$. In the lower universe, on the other hand, a particle with spin $s=0.1$ moves in a stable circular orbit at the ISCO radius $l=1.470809497$ with angular momentum $\mathcal{L}=1.997394095017$. Note that the ISCO in the lower universe has a higher angular momentum, and it is closer to the wormhole's throat. When we change the ``minus'' configuration, the roles are inverted. See the black and red dots in the figure. In the ``minus'' configuration ($s$ positive and $\mathcal{L}$ negative), the ISCO represented by the red dot is at the lower universe. 
 
	\begin{figure*}[t]
		\begin{center}$
			\begin{array}{ccc}
			\includegraphics[scale=0.3]{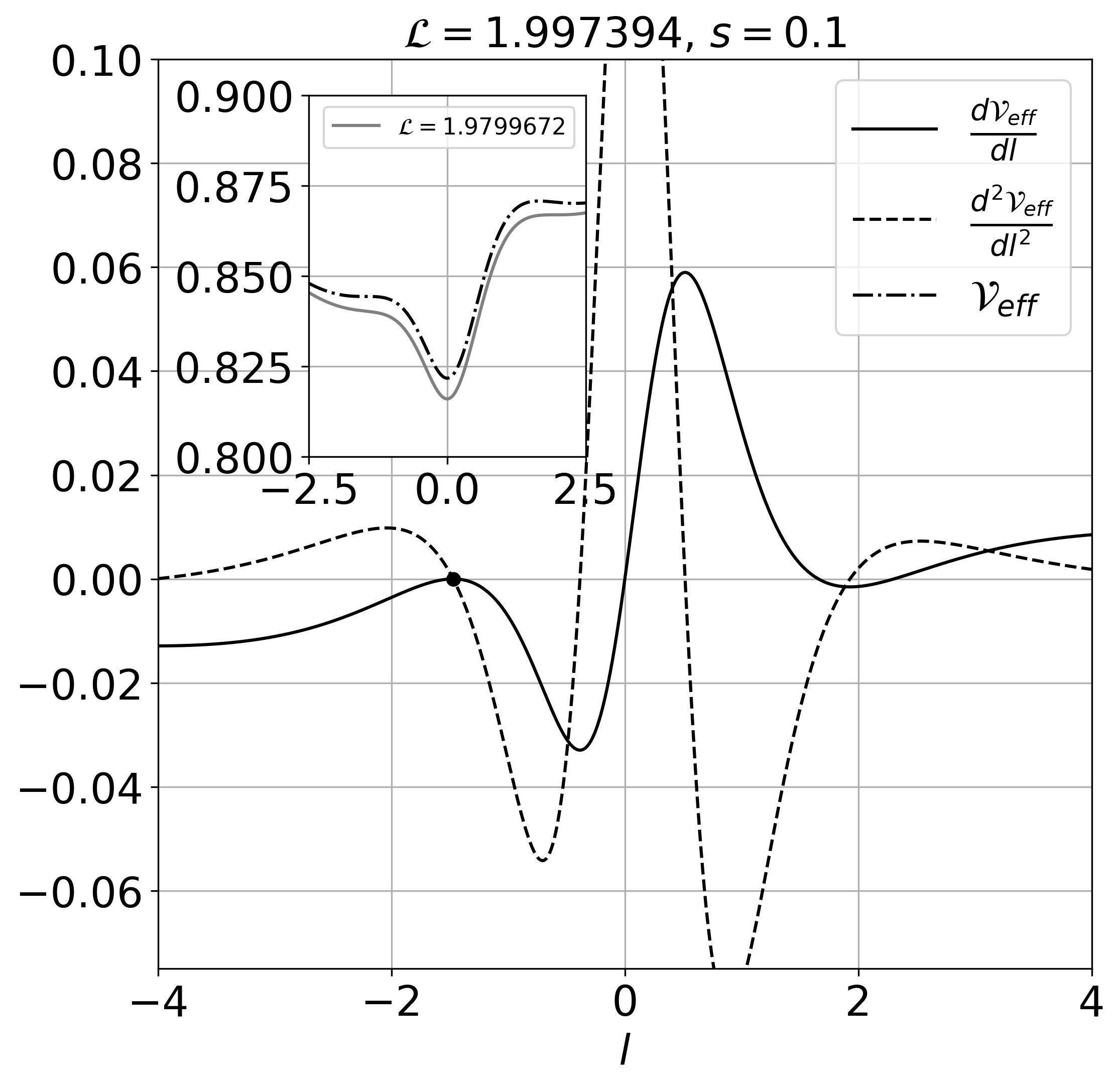} &
			\includegraphics[scale=0.3]{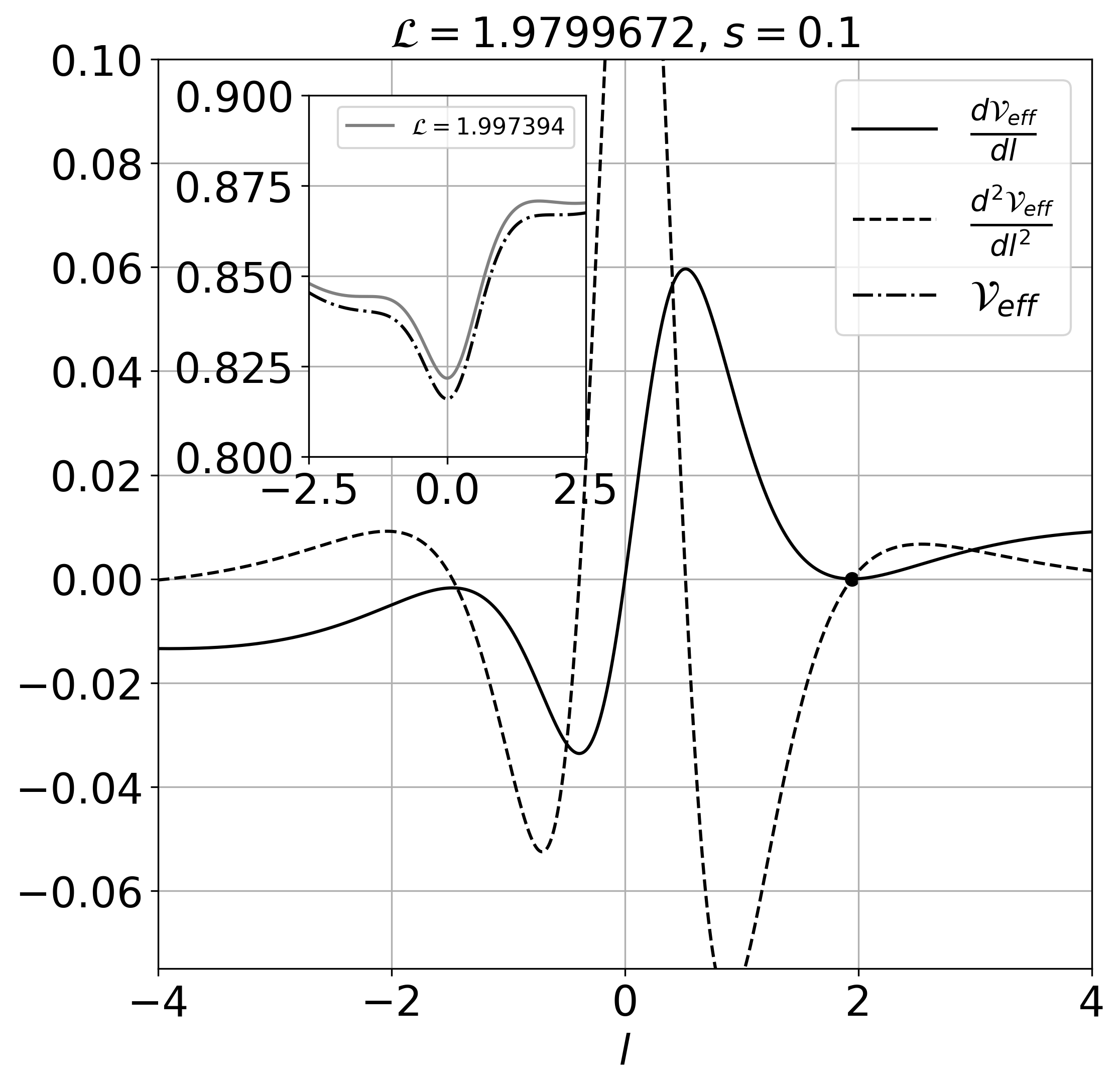}&
			\includegraphics[scale=0.3]{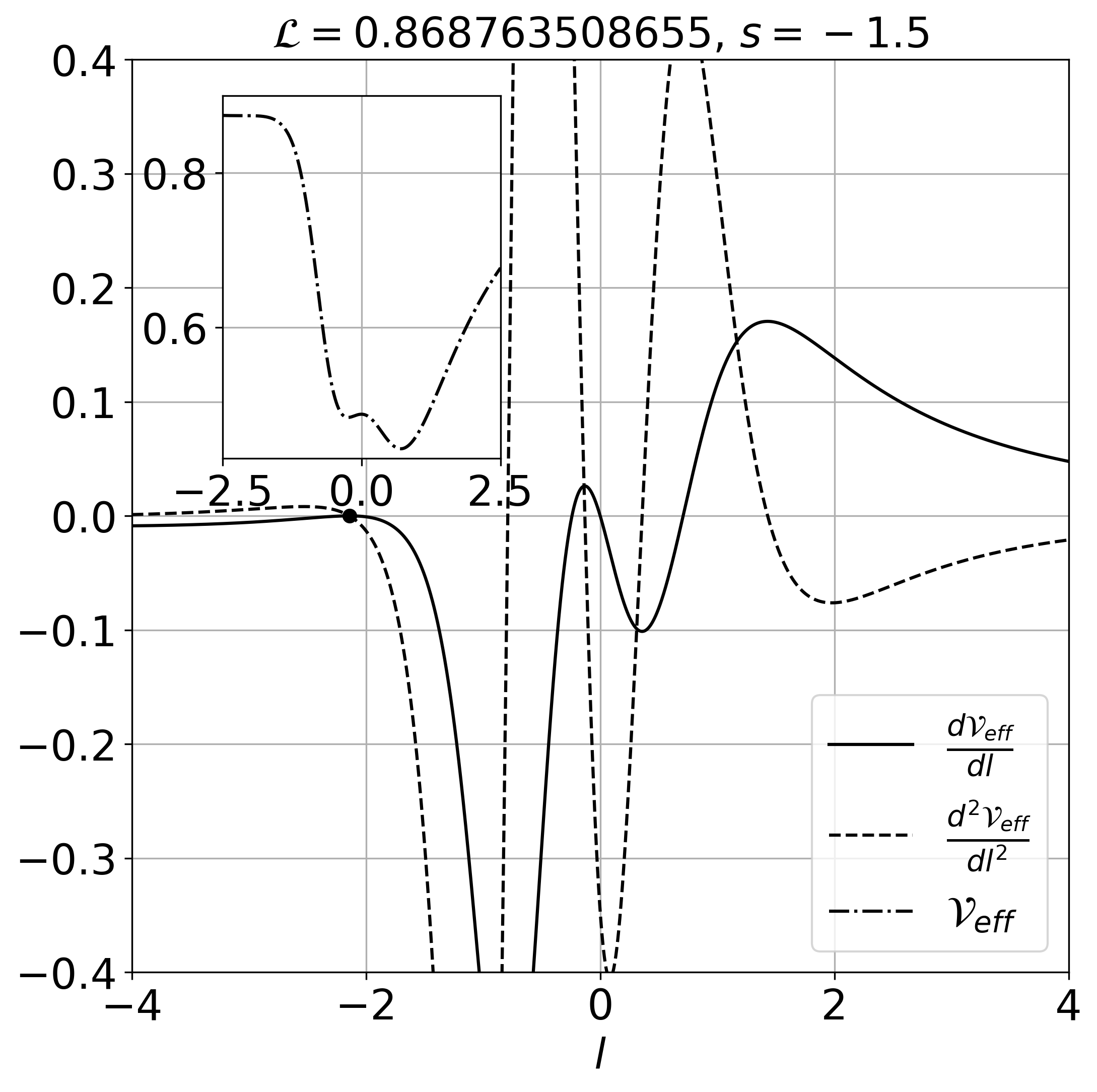}
			\end{array}$
		\end{center}
		\caption{First and second derivatives of the effective potential as a function of $l$. In the left panel with $s=0.1$ and $\mathcal{L}=1.997394095017$, in the middle panel with $s=0.1$ and $\mathcal{L}=1.979967293191$ and in the right panel, we consider the case in which $s=-1.5$ and $\mathcal{L}=0.868763508655$. In the plots we consider $b_0=M=1$, we also include, in the small frame, a plot of the effective potentials for comparison. The ISCO is given by $\mathcal{V}'_{\text{eff}}=\mathcal{V}''_{\text{eff}}=0$ and we can see that for $s=0.1$ there are two ISCO, one for $l<0$ and one for $l>0$, while for $s=-1.5$ there is only one ISCO for $l<0$.
		\label{fig7}}
	\end{figure*}
	
	\begin{table*}[t]
    \caption{\label{tab:table2}
    Values for the radius $l_{\text{ISCO}}$, the angular momentum $\mathcal{L}_{\text{ISCO}}$ and the energy $\mathcal{E}_\text{ISCO}$ of the innermost stable circular orbit for different values of the spin $s$.}
    \begin{ruledtabular}
    \begin{tabular}{cccccccc}
    $s$&$l_{\text{ISCO}}$ &$\mathcal{L}_{\text{ISCO}}$&$\mathcal{E}_{\text{ISCO}}$&$s$&$l_{\text{ISCO}}$&$\mathcal{L}_{\text{ISCO}}$&$\mathcal{E}_{\text{ISCO}}$\\
    \hline
    $\mp$ 1.6& $\mp$ 1.920360608 & 0.722911422497 & 0.8612999409591453&$\mp$ 0.7
    &$\pm$ 0.429511438& 1.402514112333 &0.6351506341677197  \\
    $\mp$ 1.5\footnotemark[3]&$\mp$ 2.141902280 & 0.868763508655 & 0.8742579899441784 &$\mp$ 0.6
    & $\mp$ 2.534233144 & 1.713278206766 & 0.8872844627862982 \\
    $\mp$ 1.4& $\mp$ 2.322036598 & 0.993562472504 & 0.8814014855529899 &$\mp$ 0.6
    & $\pm$ 0.526940168 & 1.554036117274 & 0.682349979652707 \\
    $\mp$ 1.3\footnotemark[2]& $\mp$ 2.452395852 & 1.105236494336 & 0.8855957905171136&$\mp 0.5$
    & $\mp$ 2.464002952 & 1.780911539067 & 0.8851281413233677 \\
    $\mp$ 1.2& $\mp$ 2.541002804 & 1.208099326345 &0.8881263685022468 &$\mp$ 0.5
    & $\pm$ 0.634414281 & 1.682361990518 & 0.7239081520286046 \\
    $\mp$ 1.1& $\mp$ 2.596660369 &1.304401541915 & 0.889587124437354 &$\mp$ 0.4
    &$\mp$ 2.373390705 &1.842804901024 & 0.8822631357857148 \\
    $\mp$ 0.9&$\mp$ 2.631997715 &1.481545447701 & 0.8903365076796163 &$\mp$ 0.4
    &$\pm$ 0.763940166 & 1.795084655659 & 0.7618079197491819 \\
    $\mp$ 0.9&$\pm$ 0.230465057 & 0.928262126118 & 0.50216126965439 &$\mp$ 0.2
    & $\mp$ 2.119571659& 1.944481810011 & 0.8735759021557974 \\
    $\mp$ 0.8& $\mp$ 2.618174005 & 1.563289723767 & 0.8898509886736642 &$\mp$ 0.2
    &$\pm$ 1.183708000 & 1.962437611319 & 0.0.824239661146142 \\
    $\mp$ 0.8& $\pm$ 0.334514462 & 1.209867268321 & 0.5781567283131726 &$\mp$ 0.1\footnotemark[1]
    &$\mp$ 1.946354727 & 1.979967293191 &0.8669581544832881 \\
    $\mp$ 0.7& $\mp$ 2.585463638 & 1.640598254787 & 0.8888394718973344 &$\mp$ 0.1\footnotemark[1] &$\pm$ 1.470809497 &1.997394095017 & 0.8443614729992029 \\
    $\mp$ 0.7& $\pm$ 0.429511438 & 1.402514112333 & 0.6351506341677197 &0.0 &$\mp$ 1.732050808 &1.999999999999 & 0.8577638849606256 \\
    \end{tabular}
    \end{ruledtabular}
    \footnotetext[1]{See the first row of Fig.~\ref{fig6}, and the left and central panels in Fig.~\ref{fig7}.}
    \footnotetext[2]{See the second row of Fig.~\ref{fig6}.}
    \footnotetext[3]{See the right panel of Fig.~\ref{fig7}.}
    \end{table*}
	For values of the spin in the range $-1<s<1$, we have two values for the ISCO at each universe. Nevertheless, for $s<-1$ or $s>1$, we have only one ISCO; located only in one of the universes: upper or lower. We can see this behavior from the contour plots in the second row of Fig.~\ref{fig6}, where we impose $s=1.3$. 
	\begin{figure*}[t]
		\begin{center}$
			\begin{array}{ccc}
			\includegraphics[scale=0.32]{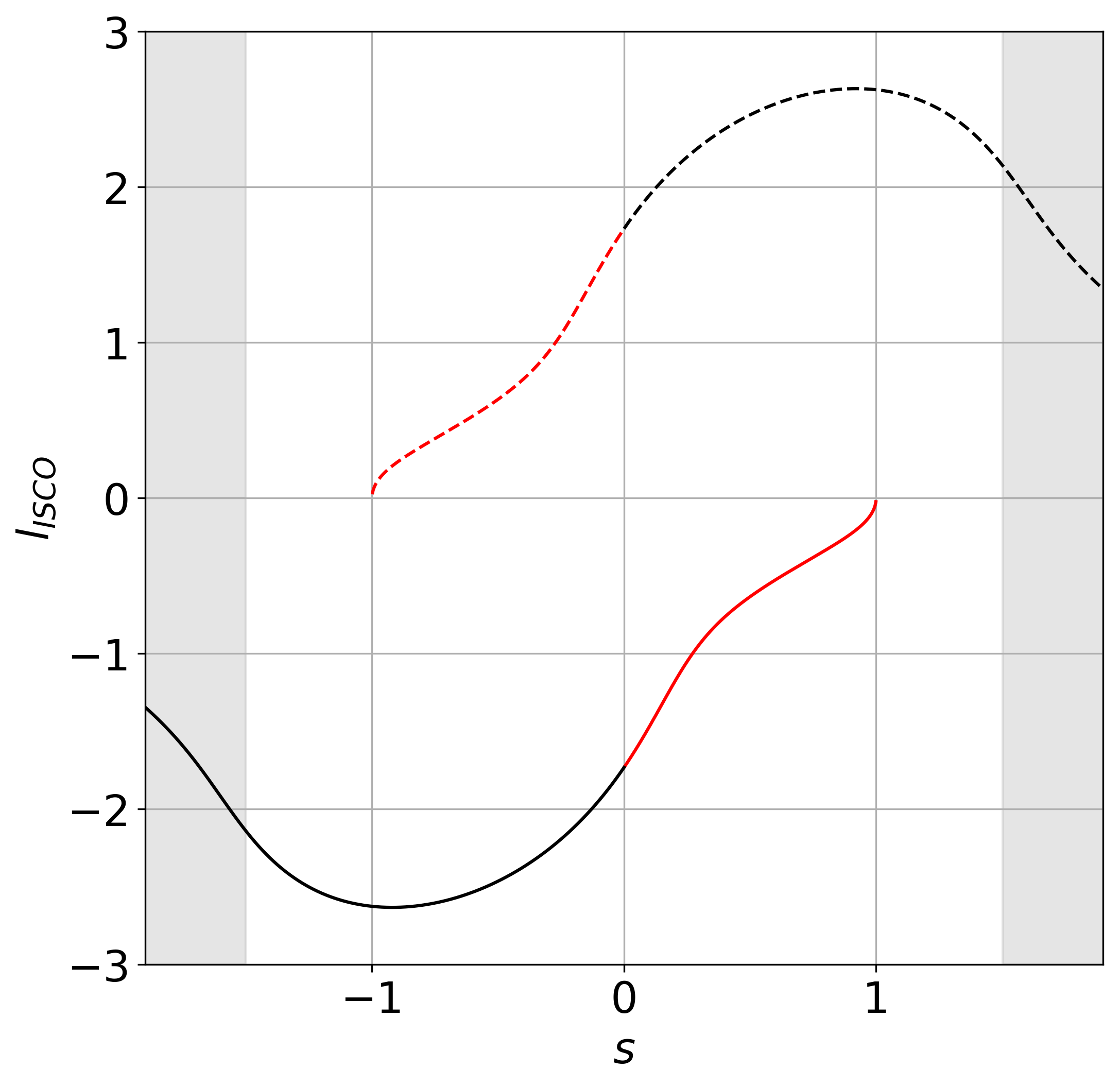} &
			\includegraphics[scale=0.32]{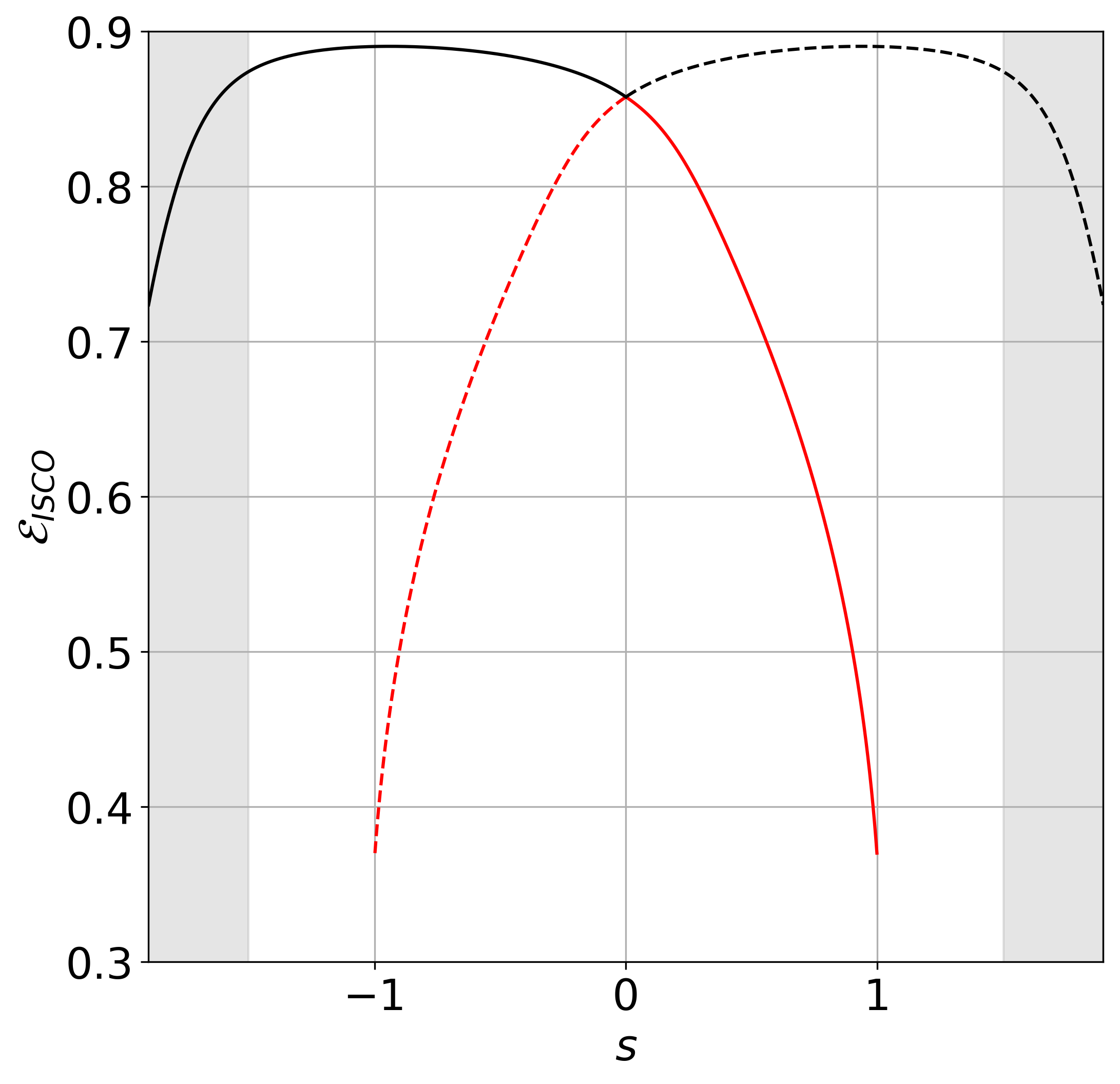}&\\
			\includegraphics[scale=0.32]{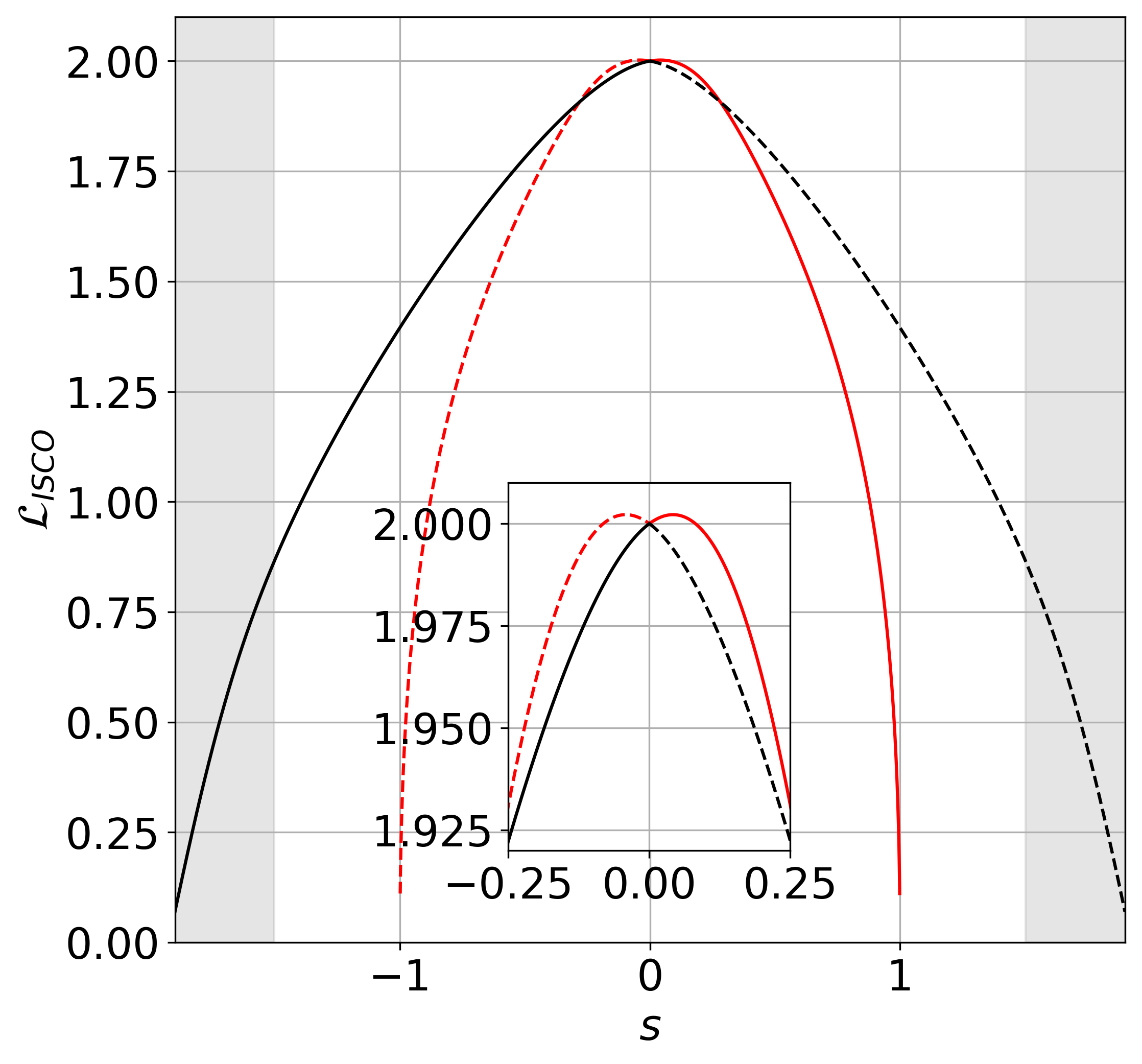}&
			\includegraphics[scale=0.32]{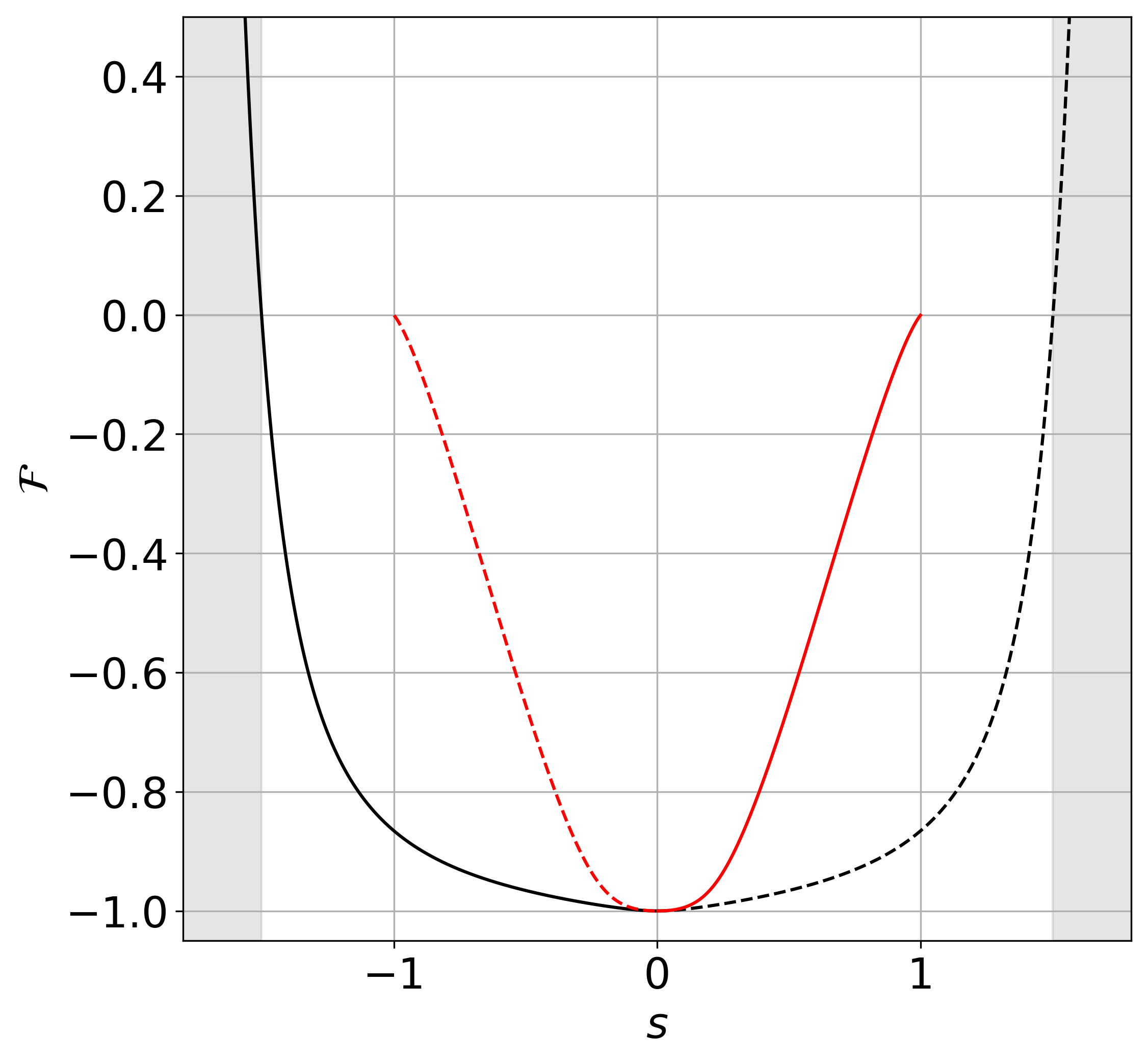}&
			\end{array}$
		\end{center}
		\caption{Plots of ISCO radius $l_{\text{ISCO}}$, energy $\mathcal{E}_{\text{ISCO}}$, angular momentum $\mathcal{L}_{\text{ISCO}}$, and $\mathcal{F}$ (see Eq.~(\ref{3.48})) as a function of $s$. The curves in red color corresponds to the values of the ISCO closer to the wormhole's throat (see table~\ref{tab:table2}). The region in gray color represents the zone where the particle's motion is space-like. In the plots we assume $b_0=M=1$.\label{fig8}}
	\end{figure*}
	
    The above statement is clearly illustrated in Fig.~\ref{fig7}. Nevertheless, taking into account the discussion of Sec.\ref{SecIIIC}, the interval $|s|>1$ does not represent realistic values of the spin. In the case of an extended body like Jupiter, for example, the ratio $S/m^2$ could be $\mathcal{O}(10^3)$, then $s\sim 1$. Therefore, although Jupiter may survive a traversable wormhole (because there is no tidal), the MPD equations are inappropriate to model its motion. Only those values of $s<<1$ have physical meaning.
	
	In the figure, we plot the first and second derivatives of the effective potential when $s=0.1$ (left and central panels) and $s=-1.5$ (right panel), with the corresponding values of the angular momentum being $\mathcal{L}_\text{ISCO}=1.979967293191$, $1.997394095017$ and $0.868763508655$, respectively.  In the left panel, we see how the derivatives intersect at $l_{\text{ISCO}}=-1.4708095$ (lower universe) when $\mathcal{L}=1.997394$. Moreover, from the small frame, we see that the values of the corresponding effective potential (Dot-dashed line) are greater than the values of $\mathcal{V}_{\text{eff}}$ when the angular momentum is $\mathcal{L}=1.979967293191$ (Gray-line). In the central panel of Fig.~\ref{fig7}, we see the intersection of the first and second derivatives at $l_{\text{ISCO}}=+1.946354727$ (upper universe). In the small frame we plot the effective potential when $\mathcal{L}=1.979967293191$ (Dot-Dashed line) and $\mathcal{L}=1.997394095017$ (gray line) for reference. In the right panel, we consider the case when $s=-1.5$ and $\mathcal{L}=0.868763508655$ and in this case, the first and second derivatives intersect at $l=-2.141902280$ (lower universe). Note that this time we considered the ``minus'' configuration, but analogue results hold for the ``plus'' configuration. Other examples are listed in table~\ref{tab:table2}.
	
	Using the values of table~\ref{tab:table2}, one can also check the superluminal bound by means of Eq.~(\ref{3.48}). Hence, for a wormhole described by Eqs.~(\ref{3.11a}) and (\ref{3.11b}), we have 
    \begin{equation}
    \begin{aligned}
    \frac{p^t}{m}&=-\frac{1}{1-\frac{l^2s^2}{\left(l^2+1\right)^{5/2}}}\left(\frac{l \sqrt{e^{-\frac{2}{\sqrt{l^2+1}}} \left(l^2+1\right)} s \mathcal{L}}{\left(l^2+1\right)^{5/2}}+\mathcal{E}\right)\\
    \frac{p^\varphi}{m}&=\frac{1}{1-\frac{l^2 s^2}{\left(l^2+1\right)^{5/2}}}\left(\frac{l s \mathcal{E}}{\sqrt{e^{-\frac{2}{\sqrt{l^2+1}}} \left(l^2+1\right)}}+\mathcal{L}\right),
    \end{aligned}
    \end{equation}
    and
    \begin{widetext}
    \begin{equation}
        \begin{aligned}
        \mathcal{X}&=e^{\frac{2}{\sqrt{1+l^2}}}\left\{\left(\frac{l^2s^2}{(1+l^2)^\frac{5}{2}}-1\right)^2-\frac{[(1+l^2)^2-s^2]^2}{(1+l^2)^4}\right\},\\\\ \mathcal{Y}&=\Bigg\{\left[(1+l^2)^\frac{7}{2}-\left(1-l^2(1+2l^2)+l^2\sqrt{1+\l^2}\right)s^2\right]^2-(1+l^2)^2
    \left[(1+l^2)^\frac{5}{2}-l^2s^2\right]\Bigg\}\frac{1}{(1+l^2)^8},\\\\
    \mathcal{Z}&=1-\frac{l^2s^2}{(1+l^2)^\frac{5}{2}}.
        \end{aligned}
    \end{equation}
    \end{widetext}
    In Fig.~\ref{fig8}, we show the behavior of $l_{ISCO}$, $\mathcal{E}_{ISCO}$, $\mathcal{L}_{ISCO}$ and $\mathcal{F}_{ISCO}$ as a function of $s$. In the first row left panel, for example,  we plot $l_{ISCO}$ as a function of $s$. As mentioned before, note there is only one value for $l_{ISCO}$ (in the upper or lower universes) when $|s|\geq1$ and two values of $l_{ISCO}$ when $-1< s<1$ (one in the upper universe and the other in the lower universe). For this reason, we have two different colors representing these intervals: black (when $l_{ISCO}$ is far from the throat) and red (for those values of $l_{ISCO}$ closer to the throat) colors. We have a similar situation for $\mathcal{E}_{ISCO}$, $\mathcal{L}_{ISCO}$ and $\mathcal{F}_{ISCO}$. In the lower universe, the figure shows that $l_{ISCO}$ decreases when the spin of the particle $s\leq-1$. Then, in the interval $-1< s < 1$, $l_{ISCO}$ increases and gets closer to the throat of the wormhole. In the upper universe, on the other hand, the situation is inverted. Note that we have chosen $\mathcal{L}_{ISCO}$ always positive so that the ``plus'' and ``minus'' configurations are given only by the sign of $s$.  

    In the first row right panel of Fig.~\ref{fig8}, we show the behavior of $\mathcal{E}_{ISCO}$ as a function of $s$. As expected, when $s=0$, the value of $\mathcal{E}_{ISCO}$ is the same in both universes due to the symmetry in the effective potential. On the other hand, when  $-1< s\leq0$ (``minus'' configuration), note that particles in the lower universe move with a higher value of energy than those moving in the upper universe, but it increases as $s$ increases reaching the same value at $s=0$. The situation is inverted when we change the configuration to ``puls'' ($ 0\leq s <1$). 

    In the second row left panel of Fig.~\ref{fig8}, we show the behavior of $\mathcal{L}_{ISCO}$ as a function of $s$. From the figure, when $-1< s\leq0$, it is possible to see that particles in the lower universe move with higher angular momentum than those in the upper universe. However, at some value of $s$ (around $-0.25$), particles in the upper universe begin to move with higher angular momentum with a maximum value $\mathcal{L}_{ISCO}=2.0022010837825808$, see the small frame added in the figure. Note that this maximum value occurs far from the throat when $l_{ISCO}=1.627928721886677$. Once again, the situation is inverted when we consider the ``plus'' configuration.
    
    Finally, in the second row right panel of Fig.~\ref{fig8}, we show the value of $\mathcal{F}_{ISCO}$ (see Eq.~(\ref{3.48})). From the figure, it is clear that the condition $\mathcal{F}<0$ is always satisfied when the particle's spin belongs to the interval $-1<s<1$. The last statement is true for particles moving in both universes. However, when $|s|>1$, the value of $\mathcal{F}$ increases and reaches the zero value at $s=\mp1.5$, which is the well-known superluminal bound. Therefore, when the spin is in the interval $-1.5<s<1.5$, the motion of a spinning particle at the ISCO is time-like and has physical meaning. Nevertheless, when $|s|>1.5$, the motion of a spinning particle at the ISCO is space-like and meaningless from the physical point of view; see the gray region in all the figures.
    
    \section{Conclusions\label{SecV}}
    
   In this work, we have studied the motion of spinning test particles around a traversable wormhole using the MPD equations. These equations relate the Riemann curvature tensor to the second rank and skew-symmetric tensor $S^{\alpha\beta}$, which is related to the particle spin $s$. Hence, following Refs.~\cite{Conde:2019juj, Toshmatov:2019bda}, we were able to compute the effective potential for a spinning test particle moving in the background of a traversable wormhole, described by the Morris-Thorne solution given in
   Refs.~\cite{Li:2014coa,Harko:2008vy}. 
   
   Our analysis shows how the effective potential depends on the angular momentum, the throat of the wormhole $b_0$, the proper radial distance $l$ and the spin of the particle $s$. In particular, we discussed the symmetries of the system and the location of the ISCO in upper and lower universes. These symmetries depend on the signs of the angular momentum and the spin of the particle. When $-1<s<1$, one interesting feature we found is the existence of two radii allowed for the ISCO on both sides of the wormhole's throat depending on whether the particle's spin is co-rotating or counter-rotating with respect to the particle's angular momentum. On the other hand, when $|s|>1$, only one ISCO exists for co-rotation in the upper universe and counter-rotation in the lower universe. This would produce some peculiar features for accretion disks around wormholes that will be investigated in future works.
  
   Finally, we studied the superluminal bound for spinning particles at the ISCO and showed that they are allowed to move in a time-like trajectory if $-1.5<s<1.5$. In this region, the value of $\mathcal{F}$ is negative, satisfying Eq.~(\ref{3.48}). Nevertheless, for values of $|s|>1.5$, the motion of the spinning test particles is space-like, and therefore meaningless from the physical point of view. On the other hand, in the case of compact objects orbiting a traversable wormhole, our analysis indicates that they move in a time-like trajectory because the spin parameter of these objects is $s<<1$. In this particular case, our analysis shows that compact objects orbiting a traversable wormhole in the lower universe will move with higher values of energy $\mathcal{E}_{ISCO}$ than those moving in the upper universe. We can see similar behavior in the case of the angular momentum $\mathcal{L}_{ISCO}$: compact objects in the lower universe move with higher angular momentum than those in the upper universe. However, at some value of the spin $s\approx -1.5$, compact objects in the upper universe begin to move with higher angular momentum.
   
  When considering astrophysical sources such as stellar mass black hole candidates or supermassive black hole candidates, then the motion of test particles orbiting the central object may allow us to determine the features that distinguish a central black hole from a wormhole~\cite{Bambi:2021qfo,Jusufi:2021lei,Tripathi:2019trz}. Although, considering the discussion of Sec.\ref{SecIIIC}, it is important to remember that the MPD equations are obtained under the assumption that the  mass and  size  of  the  spinning  test  particle  must  be  negligible  with  respect  to  the central  object’s  mass  and  must not affect  the  background geometry, the presence of the spin may still produce observable effects on the motion of test particles.
  Such test particles may be in the form of gas particles in the accretion disk but also in the form of large objects such as asteroids and planets orbiting stellar mass candidates and rapidly rotating black holes and neutron stars orbiting supermassive candidates. In this respect then, the spin of the test particle may become an important element to consider when describing the motion of such objects and the observations that would allow us to conclude if these objects are black holes or wormholes.

   \begin{acknowledgments}
    The work of C.A.B.G and A.A. is supported by a postdoc fund through PIFI of the Chinese Academy of Sciences. WH is supported by NSFC No. 11773059. A.A. and B.A. acknowledge the support of the Uzbekistan Ministry for Innovative Development.
   \end{acknowledgments}

	\appendix
	\section{Derivation of $u^l$ and $u^\varphi$\label{A1}}
    Let us start with the first relation in Eq.~(\ref{3.31}). From the Tulczyjew-SSC condition $S^{\alpha\beta}p_\alpha=0$, we obtain
    \begin{equation}
    \label{A.1}
    S^{tl}p_t+S^{\varphi l}p_\varphi=0.
    \end{equation}
    Applying the operator $D/d\lambda$, we get
    \begin{equation}
    \label{A.2}
    \begin{aligned}
    \frac{D}{d\lambda}(S^{tl}p_l)+\frac{D}{d\lambda}(S^{\varphi l}p_\varphi)&=
    p_t\frac{D S^{tl}}{d\lambda}+S^{tl}\frac{D p_t}{d\lambda}\\
    &+p_\varphi\frac{D S^{\varphi l}}{d\lambda}+S^{\varphi l}\frac{D p_\varphi}{d\lambda}=0,
    \end{aligned}
    \end{equation}
    from which, after recalling that $S^{tl}=-(p_\varphi/p_t)S^{\varphi l}$, solving for $DS^{tl}/d\lambda$, and inserting into the MPD equation, we obtain
    \begin{equation}
    \label{A.3}
    \frac{DS^{tl}}{d\lambda}=\frac{p_\varphi S^{\varphi l}}{(p_t)^2}\frac{Dp_t}{d\lambda}
    -\frac{p_\varphi}{p_t}\frac{DS^{\varphi l}}{d\lambda}-\frac{S^{\varphi l}}{p_t}\frac{Dp_\varphi}{d\lambda}=p^tu^l-u^tp^l.
    \end{equation}
    In a similar way, we proceed with $D S^{t\varphi}/d\lambda$. The Tulczyjew-SSC condition takes the form
    \begin{equation}
    \label{A.4}
    S^{\alpha\varphi}p_\alpha=S^{l\varphi}p_l+S^{t\varphi}p_t=0.
    \end{equation}
    Then, applying the operator $D/d\lambda$, we obtain
    \begin{equation}
    \label{A.5}
    \begin{aligned}
    \frac{D}{d\lambda}(S^{l\varphi}p_l)+\frac{D}{d\lambda}(S^{t\varphi}p_t)&=
    p_l\frac{D S^{l\varphi}}{d\lambda}+S^{l\varphi}\frac{Dp_l}{d\lambda}\\
    &+p_t\frac{DS^{t\varphi}}{d\lambda}+S^{t\varphi}\frac{Dp_t}{d\lambda}=0.
    \end{aligned}
    \end{equation}
    Recalling that $S^{t\varphi}=(p_l/p_t)S^{\varphi l}$, taking into account the $S^{l\varphi}=-S^{\varphi l}$, solving for $D S^{t\varphi}/d\lambda$, and inserting into the MPD equation, we obtain 
    \begin{equation}
    \label{A.6}
    \frac{DS^{t\varphi}}{d\lambda}=\frac{p_l}{p_t}\frac{DS^{\varphi l}}{d\lambda}+\frac{S^{\varphi l}}{p_t}\frac{Dp_l}{d\lambda}-\frac{p_lS^{\varphi l}}{(p_t)^2}\frac{Dp_t}{d\lambda}=p^tu^\varphi-u^t p^\varphi.
    \end{equation}
    Now, from the MPD equation for $S^{\varphi l}$, the last two of Eqs.~(\ref{3.31}) reduce to 
    \begin{equation}
    \label{A.7}
    \begin{aligned}
    \frac{S^{\varphi l}}{p_t}\left(p_\varphi\frac{Dp_t}{d\lambda}-p_t\frac{Dp_\varphi}{d\lambda}\right)&=(p_\varphi p^\varphi+p_tp^t)u^l\\
    &-p^lp_\varphi u^\varphi-p^lp_tu^t,\\
    \frac{S^{\varphi l}}{p_t}\left(p_t\frac{Dp_l}{d\lambda}-p_l\frac{Dp_t}{d\lambda}\right)&=(p_lp^l+p_tp^t)u^\varphi\\
    &-p^\varphi p_lu^l-p^\varphi p_tu^t.
    \end{aligned}
    \end{equation}
    Now we use the second MPD equation to obtain $Dp_t/d\lambda$, $Dp_\varphi/d\lambda$ and $Dp_l/d\lambda$. Hence from 
    \begin{equation}
    \label{A.8}
    \frac{Dp_\alpha}{d\lambda}=-\frac{1}{2}R_{\alpha\beta\delta\sigma}u^\beta S^{\delta\sigma}
    \end{equation}
    we obtain
    \begin{equation}
    \label{A.9}
    \frac{Dp_t}{d\lambda}=-\frac{1}{2}\left(2R_{tt\delta\sigma}u^tS^{\delta\sigma}
    +2R_{tl\delta\sigma}u^lS^{\delta\sigma}+2R_{t\varphi \delta\sigma}u^\varphi S^{\delta\sigma}\right).
    \end{equation}
    The factor 2 in the Riemann tensor components comes due to the antisymmetry of both $S^{\alpha\beta}$ and $R_{\alpha\beta\delta\sigma}$. Therefore, we have to count twice in the sum because $R_{ab\delta\sigma}S^{\delta\sigma}=R_{ab\sigma\delta}S^{\sigma\delta}$. Consequently, we get
    \begin{equation}
    \label{A.10}
    \frac{Dp_t}{d\lambda}=-R_{tltl}u^l S^{tl}-R_{t\varphi t\varphi}u^\varphi S^{t\varphi}.
    \end{equation}
    Recalling that $S^{tl}=-\frac{p_\varphi}{p_t}S^{\varphi l}$ and $S^{t\varphi}=\frac{p_l}{p_t}S^{\varphi l}$, the last equation reduces to
    \begin{equation}
    \label{A.11}
    \frac{Dp_t}{d\lambda}=\frac{S^{\phi l}}{p_t}\left(p_\varphi R_{tltl}u^l-p_l R_{t\varphi t\varphi}u^\varphi\right).
    \end{equation}
    Following a similar procedure, we obtain the following relations for $Dp_l/d\lambda$ and $Dp_\varphi/d\lambda$ 
    \begin{equation}
    \label{A.12}
    \begin{aligned}
    \frac{Dp_l}{d\lambda}&=\frac{S^{\varphi l}}{p_t}(p_\varphi R_{lttl}u^t+p_t R_{l\varphi l \varphi}u^\varphi),\\
    \frac{Dp_\varphi}{d\lambda}&=\frac{S^{\varphi l}}{p_t}(p_t R_{\varphi ll\varphi}u^l-p_l R_{\varphi tt\varphi}u^t).
    \end{aligned}
    \end{equation}
    Then, inserting them into the system of equations (\ref{A.7}), we obtain
    \begin{equation}
    \label{A.13}
    \begin{aligned}
    u^l\left[(p_\varphi)^2\hat{\mathcal{A}}+(p_t)^2\hat{\mathcal{B}}\right]&=u^\varphi\hat{\mathcal{C}}p_l p_\varphi+u^t\hat{\mathcal{C}}p_l p_t,\\
    u^\varphi\left[(p_l)^2\hat{\mathcal{C}}+(p_t)^2\hat{\mathcal{B}}\right]&=u^l\hat{\mathcal{A}}p_l p_\varphi+u^t\hat{\mathcal{A}}p_{\varphi} p_t.
    \end{aligned}
    \end{equation}
    Finally, solving the system, we obtain 
    \begin{equation}
    \label{A.14}
    \begin{aligned}
    u^l&=\frac{\hat{\mathcal{C}}}{\hat{\mathcal{B}}}\frac{p_l}{p_t},\\
    u^\varphi &=\frac{\hat{\mathcal{A}}}{\hat{\mathcal{B}}}\frac{p_\varphi}{p_t},
    \end{aligned}
    \end{equation}
    with
    \begin{equation}
    \label{A.15}
    \begin{aligned}
    \hat{\mathcal{A}}&=g^{\varphi\varphi}+R_{tllt}\left(\frac{S^{\varphi l}}{p_t}\right)^2,\\
    \hat{\mathcal{B}}&=g^{tt}+R_{\varphi ll\varphi}\left(\frac{S^{\varphi l}}{p_t}\right)^2,\\
    \hat{\mathcal{C}}&=g^{ll}+R_{\varphi tt\varphi}\left(\frac{S^{\varphi l}}{p_t}\right)^2.
    \end{aligned}
    \end{equation}
    \newpage


\begin{thebibliography}{99}
	
	\bibitem{Einstein:1935tc}
	A.~Einstein and N.~Rosen,
	Phys. Rev. \textbf{48}, 73-77 (1935)
	doi:10.1103/PhysRev.48.73
	
	\bibitem{Schwarzschild:1916uq}
	K.~Schwarzschild,
	Sitzungsber. Preuss. Akad. Wiss. Berlin (Math. Phys. ) \textbf{1916}, 189-196 (1916)
	[arXiv:physics/9905030 [physics]].
	
	\bibitem{Schwarzschild:1916ae}
	K.~Schwarzschild,
	Sitzungsber. Preuss. Akad. Wiss. Berlin (Math. Phys. ) \textbf{1916}, 424-434 (1916)
	[arXiv:physics/9912033 [physics.hist-ph]].
	
	\bibitem{G.W.Gibbons2015}
	G.~W.~Gibbons,
	Gen. Rel. Grav. \textbf{47} (2015) 71.
	
	\bibitem{Fuller:1962zza}
	R.~W.~Fuller and J.~A.~Wheeler,
	Phys. Rev. \textbf{128}, 919-929 (1962)
	doi:10.1103/PhysRev.128.919
	
	\bibitem{Misner:1957mt}
	C.~W.~Misner and J.~A.~Wheeler,
	Annals Phys. \textbf{2}, 525-603 (1957)
	doi:10.1016/0003-4916(57)90049-0
	
	\bibitem{HGEllis:1973}
	H.~G.~Ellis,
	Journal of Mathematical Physics. \textbf{14} (1): 104–118 (1973).
	
	\bibitem{Bronnikov:1973fh}
	K.~A.~Bronnikov,
	Acta Phys. Polon. B \textbf{4}, 251-266 (1973)
	
	\bibitem{Morris:1988cz}
	M.~S.~Morris and K.~S.~Thorne,
	Am. J. Phys. \textbf{56}, 395-412 (1988)
	doi:10.1119/1.15620
	
	\bibitem{Morris:1988tu}
	M.~S.~Morris, K.~S.~Thorne and U.~Yurtsever,
	Phys. Rev. Lett. \textbf{61}, 1446-1449 (1988)
	doi:10.1103/PhysRevLett.61.1446
	
	\bibitem{Visser:1989vq}
	M.~Visser,
	LA-UR-89-1008.
	
	\bibitem{Visser:1989kh}
	M.~Visser,
	Phys. Rev. D \textbf{39}, 3182-3184 (1989)
	doi:10.1103/PhysRevD.39.3182
	[arXiv:0809.0907 [gr-qc]].
	
	\bibitem{Visser:1989kg}
	M.~Visser,
	Nucl. Phys. B \textbf{328}, 203-212 (1989)
	doi:10.1016/0550-3213(89)90100-4
	[arXiv:0809.0927 [gr-qc]].
	
	\bibitem{AzregAinou:1989wr}
	M.~Azreg-Ainou and G.~Clement,
	NTH-89-9.
	
	\bibitem{Visser:1989am}
	M.~Visser,
	Phys. Lett. B \textbf{242}, 24-28 (1990)
	doi:10.1016/0370-2693(90)91588-3
	
	\bibitem{Poisson:1989zz}
	E.~Poisson and W.~Israel,
	Phys. Rev. Lett. \textbf{63}, 1663-1666 (1989)
	doi:10.1103/PhysRevLett.63.1663
	
	\bibitem{Visser:1990wj}
	M.~Visser,
	PRINT-90-0513 (WASH.U.,ST.LOUIS).
	
	\bibitem{Frolov:1990si}
	V.~P.~Frolov and I.~D.~Novikov,
	Phys. Rev. D \textbf{42}, 1057-1065 (1990)
	doi:10.1103/PhysRevD.42.1057
	
	\bibitem{Visser:1990wi}
	M.~Visser,
	Phys. Rev. D \textbf{43}, 402-409 (1991)
	doi:10.1103/PhysRevD.43.402
	
	\bibitem{Hawking:1974rv}
	S.~W.~Hawking,
	Nature \textbf{248}, 30-31 (1974)
	doi:10.1038/248030a0
	
	\bibitem{Martin-Moruno:2013}
	P. Martin-Moruno and M. Visser, 
	JHEP {\bf 1309}, 050 (2013)
	doi:10.1007/JHEP09(2013)050
	
	\bibitem{Wheeler:1955zz}
	J.~A.~Wheeler,
	Phys. Rev. \textbf{97}, 511-536 (1955)
	doi:10.1103/PhysRev.97.511
	
	\bibitem{Echeverria:1991nk}
	F.~Echeverria, G.~Klinkhammer and K.~S.~Thorne,
	Phys. Rev. D \textbf{44} (1991), 1077-1099
	doi:10.1103/PhysRevD.44.1077
	
	\bibitem{Deser:1992ts}
	S.~Deser and R.~Jackiw,
	Comments Nucl. Part. Phys. \textbf{20} (1992) no.6, 337-354
	[arXiv:hep-th/9206094 [hep-th]].
	
	\bibitem{Deser:1993jr}
	S.~Deser,
	Class. Quant. Grav. \textbf{10} (1993), S67-S73
	doi:10.1088/0264-9381/10/S/006
	
	\bibitem{Hochberg:1997wp}
	D.~Hochberg and M.~Visser,
	Phys. Rev. D \textbf{56}, 4745-4755 (1997)
	doi:10.1103/PhysRevD.56.4745
	[arXiv:gr-qc/9704082 [gr-qc]].
	
    \bibitem{Abdujabbarov:2016efm}
    A.~Abdujabbarov, B.~Juraev, B.~Ahmedov and Z.~Stuchl\'\i{}k,
    Astrophys. Space Sci. \textbf{361}, no.7, 226 (2016)
    doi:10.1007/s10509-016-2818-9
    
    \bibitem{Abdujabbarov:2009ad}
    A.~A.~Abdujabbarov and B.~J.~Ahmedov,
    Astrophys. Space Sci. \textbf{321}, 225-232 (2009)
    doi:10.1007/s10509-009-0023-9
    [arXiv:0903.0446 [gr-qc]].

	\bibitem{Moffat:1991xp}
	J.~W.~Moffat and T.~Svoboda,
	Phys. Rev. D \textbf{44} (1991), 429-432
	doi:10.1103/PhysRevD.44.429
	
	\bibitem{Carlini:1992jda}
	A.~Carlini,
	
	\bibitem{Bhawal:1992sz}
	B.~Bhawal and S.~Kar,
	Phys. Rev. D \textbf{46} (1992), 2464-2468
	doi:10.1103/PhysRevD.46.2464
	
	\bibitem{Letelier:1993cj}
	P.~S.~Letelier and A.~Wang,
	Phys. Rev. D \textbf{48} (1993), 631-646
	doi:10.1103/PhysRevD.48.631
	
	\bibitem{Vollick:1998qf}
	D.~N.~Vollick,
	Class. Quant. Grav. \textbf{16} (1999), 1599-1604
	doi:10.1088/0264-9381/16/5/309
	[arXiv:gr-qc/9806096 [gr-qc]].
	
	\bibitem{Perlick:2003vg}
	V.~Perlick,
	Phys. Rev. D \textbf{69} (2004), 064017
	doi:10.1103/PhysRevD.69.064017
	[arXiv:gr-qc/0307072 [gr-qc]].
	
	\bibitem{Perlick:2004tq}
	V.~Perlick,
	Living Rev. Rel. \textbf{7} (2004), 9
	
	\bibitem{Nandi:2006ds}
	K.~K.~Nandi, Y.~Z.~Zhang and A.~V.~Zakharov,
	Phys. Rev. D \textbf{74} (2006), 024020
	doi:10.1103/PhysRevD.74.024020
	[arXiv:gr-qc/0602062 [gr-qc]].
	
	\bibitem{Kardashev:2006nj}
	N.~S.~Kardashev, I.~D.~Novikov and A.~A.~Shatskiy,
	Int. J. Mod. Phys. D \textbf{16} (2007), 909-926
	doi:10.1142/S0218271807010481
	[arXiv:astro-ph/0610441 [astro-ph]].
	
	\bibitem{Bambi:2013jda}
	C.~Bambi,
	Phys. Rev. D \textbf{87} (2013), 084039
	doi:10.1103/PhysRevD.87.084039
	[arXiv:1303.0624 [gr-qc]].
	
	\bibitem{Tripathi:2019trz}
	A.~Tripathi, B.~Zhou, A.~B.~Abdikamalov, D.~Ayzenberg and C.~Bambi,
	Phys. Rev. D \textbf{101} (2020) no.6, 064030
	doi:10.1103/PhysRevD.101.064030
	[arXiv:1912.03868 [gr-qc]].
	
	\bibitem{Kidder:1992fr}
	L.~E.~Kidder, C.~M.~Will and A.~G.~Wiseman,
	Phys. Rev. D \textbf{47}, no.10, 4183-4187 (1993)
	doi:10.1103/PhysRevD.47.R4183
	[arXiv:gr-qc/9211025 [gr-qc]].
	
	\bibitem{Apostolatos:1994mx}
	T.~A.~Apostolatos, C.~Cutler, G.~J.~Sussman and K.~S.~Thorne,
	Phys. Rev. D \textbf{49}, 6274-6297 (1994)
	doi:10.1103/PhysRevD.49.6274
	
	\bibitem{Suzuki:1996gm}
	S.~Suzuki and K.~i.~Maeda,
	Phys. Rev. D \textbf{55}, 4848-4859 (1997)
	doi:10.1103/PhysRevD.55.4848
	[arXiv:gr-qc/9604020 [gr-qc]].
	
	\bibitem{Suzuki:1997by}
	S.~Suzuki and K.~i.~Maeda,
	Phys. Rev. D \textbf{58}, 023005 (1998)
	doi:10.1103/PhysRevD.58.023005
	[arXiv:gr-qc/9712095 [gr-qc]].
	
		
	\bibitem{Saijo:1998mn}
	M.~Saijo, K.~i.~Maeda, M.~Shibata and Y.~Mino,
	Phys. Rev. D \textbf{58}, 064005 (1998)
	doi:10.1103/PhysRevD.58.064005
	
	\bibitem{Semerak:1999qc}
	O.~Semerak,
	Mon. Not. Roy. Astron. Soc. \textbf{308} (1999), 863-875
	doi:10.1046/j.1365-8711.1999.02754.x
	
	\bibitem{Kyrian:2007zz}
	K.~Kyrian and O.~Semerak,
	Mon. Not. Roy. Astron. Soc. \textbf{382}, 1922 (2007)
	doi:10.1111/j.1365-2966.2007.12502.x
	
	\bibitem{Plyatsko:2013xza}
	R.~Plyatsko and M.~Fenyk,
	Phys. Rev. D \textbf{87}, no.4, 044019 (2013)
	doi:10.1103/PhysRevD.87.044019
	[arXiv:1303.4707 [gr-qc]].
	
	\bibitem{Hackmann:2014tga}
	E.~Hackmann, C.~L\"ammerzahl, Y.~N.~Obukhov, D.~Puetzfeld and I.~Schaffer,
	Phys. Rev. D \textbf{90}, no.6, 064035 (2014)
	doi:10.1103/PhysRevD.90.064035
	[arXiv:1408.1773 [gr-qc]].
	
	\bibitem{Nucamendi:2019qsn}
	U.~Nucamendi, R.~Becerril and P.~Sheoran,
	Eur. Phys. J. C \textbf{80}, no.1, 35 (2020)
	doi:10.1140/epjc/s10052-019-7584-8
	[arXiv:1910.00156 [gr-qc]].
	
	\bibitem{Conde:2019juj}
	C.~Conde, C.~Galvis and E.~Larra\~naga,
	Phys. Rev. D \textbf{99} (2019) no.10, 104059
	doi:10.1103/PhysRevD.99.104059
	[arXiv:1905.01323 [gr-qc]].
	
	\bibitem{Toshmatov:2019bda}
	B.~Toshmatov and D.~Malafarina,
	Phys. Rev. D \textbf{100} (2019) no.10, 104052
	doi:10.1103/PhysRevD.100.104052
	[arXiv:1910.11565 [gr-qc]].
	
	\bibitem{Boonserm:2019nqq}
	P.~Boonserm, T.~Ngampitipan, A.~Simpson and M.~Visser,
	Phys. Rev. D \textbf{101}, no.2, 024050 (2020)
	doi:10.1103/PhysRevD.101.024050
	[arXiv:1909.06755 [gr-qc]].
	
    \bibitem{Han:2016cdh}
    W.~B.~Han and R.~Cheng,
    Gen. Rel. Grav. \textbf{49}, no.3, 48 (2017)
    doi:10.1007/s10714-017-2214-y
    [arXiv:1611.07602 [gr-qc]].
    
    \bibitem{Han:2016djt}
    W.~B.~Han and S.~C.~Yang,
    Int. J. Mod. Phys. D \textbf{27}, no.01, 1750179 (2017)
    doi:10.1142/S0218271817501796
    [arXiv:1610.01534 [gr-qc]].
    
    \bibitem{Toshmatov:2020wky}
    B.~Toshmatov, O.~Rahimov, B.~Ahmedov and D.~Malafarina,
    Eur. Phys. J. C \textbf{80}, no.7, 675 (2020)
    doi:10.1140/epjc/s10052-020-8254-6
    [arXiv:2003.09227 [gr-qc]].
	
	\bibitem{Mathisson:1937zz}
	M.~Mathisson,
	Acta Phys. Polon. \textbf{6}, 163-2900 (1937)

	\bibitem{Papapetrou:1951pa}
	A.~Papapetrou,
	Proc. Roy. Soc. Lond. A \textbf{209}, 248-258 (1951)
	doi:10.1098/rspa.1951.0200
	
	\bibitem{Corinaldesi:1951pb}
	E.~Corinaldesi and A.~Papapetrou,
	Proc. Roy. Soc. Lond. A \textbf{209}, 259-268 (1951)
	doi:10.1098/rspa.1951.0201
	
		
	\bibitem{tulczyjew1959motion}
	W.~Tulczyjew,
	Acta Phys. Pol\textbf{18}, 393 (1959).
	
	\bibitem{BWTulzcyjew1962}
	B.~Tulzcyjew and W.~Tulzcyjew,
	``Recent Developments in General Relativity.'' (1962)
	
	\bibitem{moller1949definition}
	M{\o}ller, Christian,
	``On the definition of the centre of gravity of an arbitrary closed system in the theory of relativity,'' (1949)
	
	\bibitem{beiglbock1967center}
	W.~Beiglb{\"o}ck,
	Communications in Mathematical Physics \textbf{5}, 106-130 (1967)
	
	\bibitem{dixon1964covariant}
	W.~G.~Dixon,
	Il Nuovo Cimento \textbf{34}, 317-339 (1964)

	
	\bibitem{Dixon:1970zza}
	W.~G.~Dixon,
	Proc. Roy. Soc. Lond. A \textbf{314}, 499-527 (1970)
	doi:10.1098/rspa.1970.0020
	
	\bibitem{Dixon:1970zz}
	W.~G.~Dixon,
	Proc. Roy. Soc. Lond. A \textbf{319}, 509-547 (1970)
	doi:10.1098/rspa.1970.0191
	
	\bibitem{ehlers1977dynamics}
	J.~Ehlers , E.~Rudolph,
	Gen. Relat. Gravit.  \textbf{8}, 197-217 (1977)
	https://doi.org/10.1007/BF00763547

	
	
	\bibitem{Chandrasekhar:1998}
	S. Chandrasekhar
	``The mathematical theory of black holes.''
	Oxford University Press, New York, 1998.
	
	\bibitem{stationary-wormhole}
	E. Teo, Phys. Rev. D 58, 024014 (1998).
	
	\bibitem{axial-wormhole1}
	G. W. Gibbons and M. S. Volkov, 
	JCAP 05, 039 (2017),

    \bibitem{axial-wormhole2}
    G. Clement, 
    Gen. Rel. Grav. 16, 477 (1984).

    \bibitem{axial-wormhole3}
    B. Narzilloev, D. Malafarina, A. Abdujabbarov, B. Ahmedov and C. Bambi,
    arxiv: 2105.09174
    

    \bibitem{Martin-Moruno:2013sfa}
    P.~Mart\'\i{}n-Moruno and M.~Visser,
    Phys. Rev. D \textbf{88}, no.6, 061701 (2013)
    doi:10.1103/PhysRevD.88.061701
    [arXiv:1305.1993 [gr-qc]].
	
	
    \bibitem{Li:2014coa}
    Z.~Li and C.~Bambi,
    Phys. Rev. D \textbf{90}, 024071 (2014)
    doi:10.1103/PhysRevD.90.024071
    [arXiv:1405.1883 [gr-qc]].
	
    \bibitem{Harko:2008vy}
    T.~Harko, Z.~Kovacs and F.~S.~N.~Lobo,
    Phys. Rev. D \textbf{78}, 084005 (2008)
    doi:10.1103/PhysRevD.78.084005
    [arXiv:0808.3306 [gr-qc]].
    
	\bibitem{Hojman:2012me}
	S.~A.~Hojman and F.~A.~Asenjo,
	Class. Quant. Grav. \textbf{30}, 025008 (2013)
	doi:10.1088/0264-9381/30/2/025008
	[arXiv:1203.5008 [physics.gen-ph]].
	
    \bibitem{Hartl:2002ig}
    M.~D.~Hartl,
    Phys. Rev. D \textbf{67}, 024005 (2003)
    doi:10.1103/PhysRevD.67.024005
    [arXiv:gr-qc/0210042 [gr-qc]].
    
    \bibitem{Cook:1993qr}
    G.~B.~Cook, S.~L.~Shapiro and S.~A.~Teukolsky,
    Astrophys. J. \textbf{424}, 823 (1994)
    doi:10.1086/173934
    
    \bibitem{Lai:1993pa}
    D.~Lai, F.~A.~Rasio and S.~L.~Shapiro,
    Astrophys. J. \textbf{420}, 811-829 (1994)
    doi:10.1086/173606
    [arXiv:astro-ph/9304027 [astro-ph]].
	
	
	\bibitem{Harko:2009xf}
	T.~Harko, Z.~Kovacs and F.~S.~N.~Lobo,
	Phys. Rev. D \textbf{79}, 064001 (2009)
	doi:10.1103/PhysRevD.79.064001
	[arXiv:0901.3926 [gr-qc]].
	
	
    \bibitem{Bambi:2021qfo}
    C.~Bambi and D.~Stojkovic,
    Universe \textbf{7}, no.5, 136 (2021)
    doi:10.3390/universe7050136
    [arXiv:2105.00881 [gr-qc]].
    
    \bibitem{Jusufi:2021lei}
    K.~Jusufi, S.~K., M.~Azreg-A\"\i{}nou, M.~Jamil, Q.~Wu and C.~Bambi,
    [arXiv:2106.08070 [gr-qc]].
    
	\end{thebibliography}
	\end{document}